%% file: paper.tex
\documentclass[fleqn,usenatbib]{mnras}

\usepackage[T1]{fontenc}
\usepackage{ae,aecompl}

\usepackage{rotating}
\usepackage{graphicx}	
\usepackage{amsmath}	
\usepackage{amssymb}	
\usepackage{booktabs}
\usepackage{natbib}
\usepackage{textcomp}
\usepackage{xfrac}
\usepackage{afterpage}
\usepackage[normalem]{ulem}
\usepackage{threeparttable} 

\input{code}


\input{nuclides}

\input{concepts}
\input{units}
\input{vectors}
\input{formatting}
\input{symbols}
\usepackage[usenames,dvipsnames]{color}
\usepackage[]{color}

\newcommand{\cldfluid}{fluid above the convection zone}
\newcommand{\airfluid}{convective fluid}
\newcommand{\ppmstar}{\code{PPMstar}}

\title[3D1D hydro-nucleosynthesis simulations of RAWDs]{3D1D
  hydro-nucleosynthesis simulations. I. Advective-reactive
  post-processing method and its application to H ingestion into He-shell
  flash convection in rapidly accreting white dwarfs}
\author[D. Stephens et al.]{David Stephens,$^{1}$
Falk Herwig,$^{1,2,\dagger}$\thanks{E-mail: fherwig@uvic.ca}
Paul Woodward,$^{2,3}$
Pavel Denissenkov,$^{1,2,\dagger}$
\newauthor
Robert Andrassy$^{4,1}$ 
and Huaqing Mao$^{3}$\\
$^{1}$Department of Physics and Astronomy, University of Victoria, Victoria, BC, V8W 2Y2, Canada\\
$^{2}$Joint Institute for Nuclear Astrophysics - Center for the Evolution of the Elements, USA\\
$^{3}$LCSE and Department of Astronomy, University of Minnesota, Minneapolis, MN 55455, USA\\
$^{4}$Heidelberg Institute for Theoretical Studies, Schloss-Wolfsbrunnenweg 35, D-69118 Heidelberg, Germany\\
$^\dagger$NuGrid Collaboration, \href{http://nugridstars.org}{http://nugridstars.org}\\}

\date{Accepted XXX. Received YYY; in original form ZZZ}

\pubyear{2020}

\begin{document}
\label{firstpage}
\pagerange{\pageref{firstpage}--\pageref{lastpage}}
\maketitle

\begin{abstract}
We present two mixing models for post-processing of 3D hydrodynamic
simulations applied to convective-reactive \ipr\ nucleosynthesis in a
rapidly accreting white dwarf (RAWD) with $\mathrm{[Fe/H]} = -2.6$, in
which H is ingested into a convective He shell. A 1D
advective two-stream model adopts physically motivated
radial and horizontal mixing coefficients constrained by 3D
hydrodynamic simulations. A simpler approach uses diffusion
coefficients calculated from the same simulations. All 3D simulations
include the energy feedback of the
\nuclei{12}{C}$(\pt,\gamma)$\nuclei{13}{N} reaction from the
H entrainment. Global oscillations of shell H
ingestion in two of the RAWD simulations cause bursts of
entrainment of H and non-radial hydrodynamic feedback. With the same
nuclear network as in the 3D simulations, the 1D advective two-stream
model reproduces the rate and location of the H burning within the He
shell closely matching the 3D simulation predictions, as well as
qualitatively displaying the asymmetry of the $X_{\mathrm{H}}$
profiles between the up- and downstream. With a full \ipr\ network
the advective mixing model captures the difference in the n-capture
nucleosynthesis in the up- and downstream. For example,
\nuclei{89}{Kr} and \nuclei{90}{Kr} with half-lives of
\unit{3.18}{\minute} and \unit{32.3}{\second} differ by a factor 2 - 10 in the
two streams. In this particular application the diffusion approach
provides globally the same abundance distribution as the advective
two-stream mixing model. The resulting \ipr\ yields are in excellent agreement with observations of
the exemplary CEMP-r/s star CS31062-050.
\end{abstract}




\section{Introduction}

The details of mixing in convection zones within the stellar interior
modifies the evolution of stars from the main sequence to the final
white dwarf stage or supernova core collapse. The continual mixing of
H into the convective cores of intermediate and massive stars sets a
timescale for the eventual main-sequence turn off. This can be
extended from convective boundary mixing which introduces additional H
fuel to the core. Through this additional mixing of fuel into
convection zones convective boundary mixing has a cumulative effect on
the nuclear-time scale evolution of individual convection zones
including on the mixture and amount of processed elements, the size of
the convection zones and thereby on subsequent evolutionary phases
\citep[for
  example][]{herwig:00c,Young:2005eg,Denissenkov:2012cu,Battino:2016bna,Davis2019,Wagstaff:2020ce}.

Over the nuclear timescale of the H-burning during the main sequence,
the mixing of chemical species is computed using either an
instantaneous or diffusive mixing approximation. The nuclear burning
timescale is usually significantly longer than the mixing timescale
within the convective core leading to the details of the mixing within
the convective zone being sufficiently well modeled with a space-time
average theory. Such a theory, the mixing length theory
\citep[MLT][]{cox:68}, describes the energy transport and mixing
properties of a convective region using averaged quantities over space
and time.  In advanced stages of stellar evolution the nuclear burning
timescales approach convective timescales, and so the details of the
mixing become increasingly important for their structure
\citep[][]{Herwig2011,Collins:2018cp,Cote:2020bo}.  To quantify this
phenomenon, it is useful to define the Damk\"ohler number,
$\mathrm{Da} = \tau_{\mathrm{mix}} / \tau_{\mathrm{nuclear}}$, which
is the ratio of the mixing and nuclear burning timescales. As this
number becomes larger and closer to 1, the details of the mixing have
a greater impact on the burning.

3D hydrodynamic simulations are becoming increasingly accurate in
capturing the process of entrainment of fuel into a convection zone
\citep{meakin:07b,Mocak:2011jk,Woodward:2013uf} as well as the
hydrodynamic feedback resulting from the dynamic nuclear burning of
the ingested fuel
\citep{meakin:06a,Herwig:2014cx,Muller:2016kta,Yoshida:2019be,Andrassy:19,Yadav:2020gh}. Such
simulations predict how species are mixed into the convection zone and
can determine the convective-reactive nucleosynthesis for a limited
number of species without needing a post-processing.  However,
modeling any complicated nucleosynthesis, for example the
\iprn\ \citep{herwig:11}, or the detailed nucleosynthesis in the
merger of a O and a C convection shell \citep{Ritter2018} may require
networks with hundreds or thousands of species that can interact with
each other, which is well beyond the capabilities of any 3D
hydrodynamic code running on modern computing clusters. Therefore, 1D
mixing models are still required to determine any complicated
nucleosynthesis while 3D hydrodynamic simulations can be used to
determine the mixing properties of the convection zone and how the
hydrodynamic instabilities lead to the entrainment of stably
stratified material. For example, the 3D hydrodynamic simulations of
\citet{Ritter2018} modeled a convective O shell with a stable C shell
from a $15\,M_{\odot}$ stellar model of \citet{RitterSE} to obtain
estimates for the expected entrainment rate of the C-rich
material. Using diffusive mixing constrained by the mixing properties
of the convection zone and appropriate entrainment rates, the 1D
large-network nucleosynthetic post-processing models produced
significant amounts of odd-Z elements like P, Cl, K and Sc, which
could explain the underproduction of these elements in current GCE
models~\citep{Ritter2018}.  The nuclear network used in the 3D
simulations only included the energy generation from the
\nuclei{12}{C}$(\nuclei{12}{C},\alpha)$\nuclei{20}{Ne} and subsequent
\nuclei{16}{O}$(\alpha,\gamma)$\nuclei{20}{Ne} reaction.

An interesting case of convective-reactive nucleosynthesis occurs when
H is ingested into a He burning convective shell.  This triggers the
\nuclei{12}{C}$(\pt,\gamma)$\nuclei{13}{N} reaction and, after the
\nuclei{13}{N} beta decays to \nuclei{13}{C}, the
\nuclei{13}{C}$(\alpha,n)$\nuclei{16}{O} reaction can release neutrons
if the temperatures within the He shell are high enough. Sakurai's
object (V4334 Sagittarii) has a unique surface chemical composition
\citep{asplund:99a} that can be explained with H being ingested into
the He-shell flash convection zone of this post-AGB
star~\citep{herwig:11}. This results in the above chain of reactions
and produces neutron densities high enough to be in the \ipr~regime
($N_{n} \approx 10^{12} - 10^{16}$\,cm$^{-3}$) \citep{cowan:77}. The
energy generation from this H ingestion can be significant enough to
cause a split of the He convective zone in 1D stellar evolution
models. To study the nucleosynthesis in Sakurai's object,
\citet{herwig:11} used a spherically symmetric diffusive mixing model
based on MLT convective velocities, even though in those conditions
several MLT assumptions are not satisfied. This is shown explicitly in the
3D hydrodynamic simulations of this very energetic event which bring
about a global oscillation of shell H ingestion, that causes
large-scale, non-radial and fast flows \citep{Herwig:2014cx}.

A more quasi-static convective-reactive case of H-ingestion into a He
shell has been found in the models of rapidly accreting white dwarfs
(RAWDs) by \cite{Denissenkov:2017ba}. The 3D hydrodynamic simulations
of \citet{Denissenkov:19} quantified the H-entrainment rates, however
the diffusive mixing used in the post-processing was taken directly
from the 1D stellar evolution models. The neutron densities in RAWDs
reach $N_{n} \approx 10^{14}$\,cm$^{-3}$ resulting in
\ipr~nucleosynthesis.  The RAWD heavy element production could result
in a significant contribution of Kr, Rb, Sr, Y, Zr, Nb and Mo to the
solar composition \citep{Cote2018}. The convective-reactive flows in
RAWDs are fed by the \nuclei{12}{C}$(\pt,\gamma)$\nuclei{13}{N} energy
generation that comprises only $2-3\%$ of the total luminosity within
the He shell which does not lead to a global oscillation of shell H
ingestion \citep{Denissenkov:19}.

A solution to the concerns about the validity of diffusive mixing in a
convective-reactive environment and the inability of 3D hydrodynamic
simulations to simultaneously perform complex nucleosynthesis
computations is to use a 1D advective mixing model.  One such model
has been adopted for use in the post-processing nucleosynthesis of
stellar evolution models from the Monash group, \code{MONSOON}
\citep{cannon1993, Henkel2017}. The model contains two adjacent
streams of fluid flow \citep[see our \Fig{elevator} and Fig.~1
  in][]{Henkel2017}, one with fluid moving upwards and another with
fluid moving downwards. The two streams have an enforced horizontal
mixing in order to conserve mass but it can also add in additional
horizontal mixing. The radial transport velocities as well as the
additional horizontal mixing within a convection zone are estimated
with MLT.

A limitation in the two-stream methodology of \code{MONSOON} arises in
the treatment of the additional horizontal mixing. The strength of
this mixing has no dependence on the structure of the flow within the
convection zone. Is the mixing between the two streams that represent
the dominant dipolar flows of core convection the same as if the flow
field was at smaller angular scales like in shell
convection~\citep{chandrabook1961}? There is a dependence on the
horizontal mixing based on where a cell is within the convection zone
but shouldn't the mixing between the two streams, in low \mach~number
flows, be stronger near the convective boundaries where the fluid is
forced to overturn well before the convective
boundaries~\citep{Jones:2017kc}? These limitations are addressed in
the advective mixing model of this work.

In this paper, we describe our adaptation of an advective two-stream
model for post-processing of detailed 3D hydrodynamic simulations. We
create 3D hydrodynamic simulations of a RAWD model
from~\citet{Denissenkov:19} to quantify the mixing of H into the He
shell and to simulate the global flow including the energy feedback
from nuclear burning of entrained material. After extracting the
mixing information from the 3D hydrodynamic simulations for the 1D
diffusive ~\citep{Jones:2017kc} and advective mixing models, the time
evolution of the H burning in both of them is calculated.  The details
of the 3D hydrodynamic simulations, as well as the diffusive and
advective mixing approaches are outlined in \Section{allmethods}.
\Section{results} discusses the flow properties of the simulations and shows the post-processing of the 3D hydrodynamic
simulations with the diffusive and advective mixing routines.
\Section{conclusions} describes the implications of such results and
further applications.


\section{Methods}
\label{sec:allmethods}

\subsection{PPMstar simulations}
\label{sec:ppmsim}

\begin{table*}
	\centering
	\caption{Summary of the \ppmstar{} simulations that were computed for this
          work. The entrainment rates are the slopes of the linear fits shown in
          \Fig{entrain}, while the different definitions of the convective boundary are
          discussed in detail in \Sections{convbound}{modelConstraints}.}
	\label{tab:ppmstarmodels}
  \begin{threeparttable}
	\begin{tabular}{ccccccccc} 
		\toprule
		Run ID & Grid & $t_{\mathrm{sim}} \unitspace (\mathrm{min})$ & $\langle \tau_{\mathrm{conv}}^{i} \unitspace (\mathrm{min}) \rangle$ & $L_{\mathrm{He}} \unitspace (L_{\odot})$ & $r_{\mathrm{b}, \mathrm{SC}}^{ii} \unitspace (\mathrm{Mm})$ & $r_{\mathrm{b}, v_{\perp}}^{iii} \unitspace (\mathrm{Mm})$ & $\sigma_{r_{\mathrm{b}}, v_{\perp}} \unitspace (\mathrm{Mm})$ & $\dot{M}_{\mathrm{e}}^{iv} \unitspace (M_{\odot} \unitspace \mathrm{s}^{-1})$\\
		\midrule
		N15 & $768^{3}$ & 1634 & 19 & $1.46 \times 10^{8}$ & $23.86$ & $23.43$ & $0.48$ & $1.07 \times 10^{-11}$  \\
		N16 & $1536^{3}$ & 744 & 18 & $1.46 \times 10^{8}$ & $23.84$ & $23.56$ & $0.46$ & $7.21 \times 10^{-12}$  \\
		N17 & $1152^{3}$ & 631 & 9 & $14.6 \times 10^{8}$ & $26.74$ & $26.69$ & $1.19$ & $1.08 \times 10^{-10}$  \\
		\bottomrule
	\end{tabular}
  \begin{tablenotes}
    \small
  \item \textit{Notes}: $^{i}$ The average convective turn over time during the
    quasi-static phase of each run ($46 < t < 300 \unitspace \mathrm{min}$); $^{ii}$ The
    initial Schwarzschild boundary as followed in the Lagrangian
    coordinates at $t = \unit{299}{min}$; $^{iii}$ Boundary where $\partial v_{\perp} / \partial r$ has a
    minimum at $t = \unit{299}{min}$; $^{iv}$ Entrainment rate of the H-rich 
    \cldfluid{} 
  \end{tablenotes}
  \end{threeparttable}                
  
\end{table*}

The advective and diffusive post-processing methods introduced here are applied
to 3D hydrodynamic simulations of He-shell flash convection in a rapidly
accreting white dwarf \citep{Denissenkov:2017ba}. The initial stratification has
been taken from the stellar evolution model G with the metallicity [Fe/H]\,$= -2.6$
from \cite{Denissenkov:19}.

As in previous work \citep{Herwig:2014cx,Jones:2017kc} we use the \ppmstar{}
code of \citet{woodward15} with additional details provided by
\citet{Andrassy:19}. The explicit Cartesian grid code is based on the
Piecewise-Parabolic Method \citep[PPM;][]{woodward_colella81,
woodward_colella84, colella_woodward84, woodward86, woodward07}, and tracks the
advection of concentrations in a two-fluid scheme using the Piecewise-Parabolic
Boltzmann method \citep[PPB;][]{woodward86,woodward15}.

The luminosity from the \nuclei{4}{He} burning within the convection zone is
modeled with a constant volume heating. The entrained H reacts with the abundant
\nuclei{12}{C} from the triple-$\alpha$ via the
\nuclei{12}{C}$(\pt,\gamma)$\nuclei{13}{N} reaction. This reaction rate is
computed using the analytic form of the rate from \citet{angulo:99} with no
screening factor. We ignore the subsequent beta decay of \nuclei{13}{N} leaving
a total energy release per reaction of $Q = \unit{1.943}{MeV}$. The stably
stratified \cldfluid{} contains 89.4\% by
number of H and the \airfluid{} contains 14.3\% by
number of \nuclei{12}{C}. The \ppmstar{} simulations done in this work are summarized
in \Tab{ppmstarmodels}.

\subsection{Diffusive mixing model}
\label{sec:Dmethod}

\citet{Jones:2017kc} inverted the diffusion equation to derive a
radius-dependent diffusion coefficient, which produces in 1D the same
redistribution of species over the time frame of analysis as that given by their
spherically-averaged 3D simulations of O-shell convection. They measured the
rate of change in the radius-dependent mass fraction $X(r,\ t)$ of a species by
computing the difference between the mass fraction profiles at two different
points in time with some time averaging applied around them. With $\partial X /
\partial r$ also known from the spherical averages, \citet{Jones:2017kc} solve
for the unknown diffusion coefficient, which we call $D_{\mathrm{FV}}(r)$, in
the 1D Eulerian diffusion equation
\begin{equation*}
\frac{\partial X}{\partial t} = \frac{\partial}{\partial x}\left( D_{\mathrm{FV}}(r)
\frac{\partial X(r)}{\partial x} \right),
\end{equation*}
where they set $x = r$. We have improved upon this method by mapping
the results of the input 3D Eulerian simulations to a mass coordinate
$m(r)$ and inverting the Lagrangian diffusion equation
\begin{equation}
\frac{\partial X}{\partial t} = \frac{\partial}{\partial m}\left( \sigma(m)
\frac{\partial X(m)}{\partial m} \right) + \frac{\dot{q}}{\rho},
\label{eq:diff-inv-burn}
\end{equation}
where $\sigma = (4\pi r^2 \rho)^2 D_\mathrm{FV}(r)$ is the Lagrangian diffusion
coefficient, and $\dot{q}(m,\ t)$ is the destruction rate of the species by
nuclear reactions. This new approach has the following advantages. It removes
the effect of thermal expansion and contraction from $\partial X / \partial t$.
This effect has negligible influence on the results of~\citet{Jones:2017kc}, but
it is essential to account for in the expanding convection zone in these RAWD
simulations. It also properly takes into account the spherical geometry of the
problem with the radial dependence of the density and species. Included in our
RAWD simulations but not in the O-shell convection simulations of
\citet{Jones:2017kc} are nuclear reactions between species of the two distinct
fluids containing H and \nuclei{12}{C}.

In order to invert \eq{diff-inv-burn} we need to take the difference of two
averages of the fractional volume profiles, or -- in this case -- the mass
fraction profile $X(r,\ t)$. For that we estimate the time in which it takes to
diffuse across one mixing length:
\begin{equation*}
\Delta t = \frac{(\alpha_\mathrm{MLT} H_\mathrm{P})^2}{4D_{\mathrm{FV}}}
\end{equation*}
where we take $H_\mathrm{P}$ and $D_{\mathrm{FV}}$ at the radius where
$D_{\mathrm{FV}}$ is at its maximum. This diffusive timescale is $\Delta t
\approx \unit{2.5}{\minute}$.

\ppmstar{} simulations output a \textit{dump} every several thousand time steps.
For example in N16 every $\approx 2121$ time steps, where each time step is
$\approx \unit{\natlog{1.3}{-2}}{\second}$, a single \textit{dump} is output.
For each \textit{dump} at time $t_\mathrm{D}$, $D_\mathrm{FV}(r)$ is determined
based on two profiles, $X(r,\ t_\mathrm{D}\pm \Delta t)$, each being averaged
over $\Delta t$.

\begin{figure}
  \centering
	\includegraphics[width=\columnwidth]{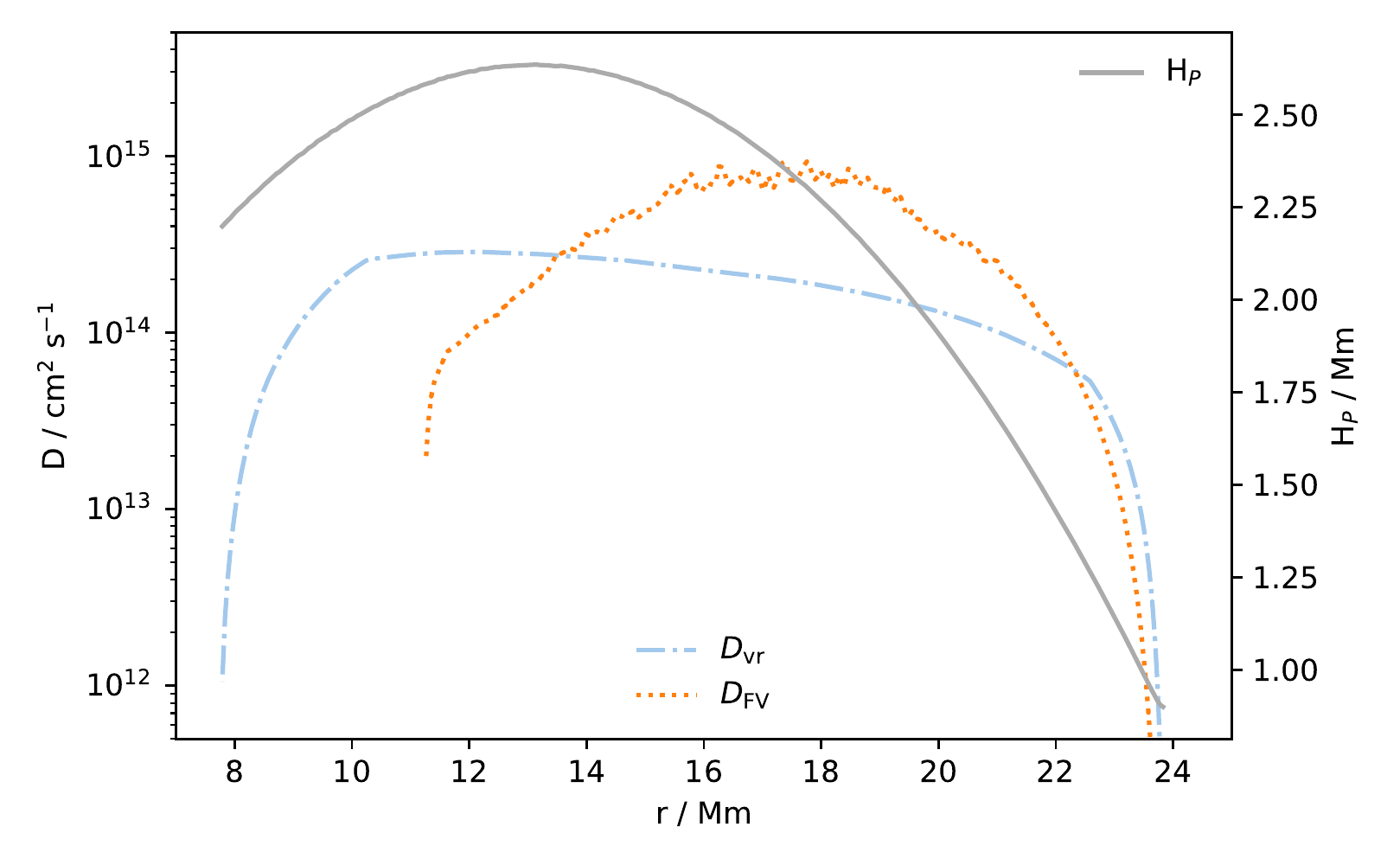}
	\includegraphics[width=\columnwidth]{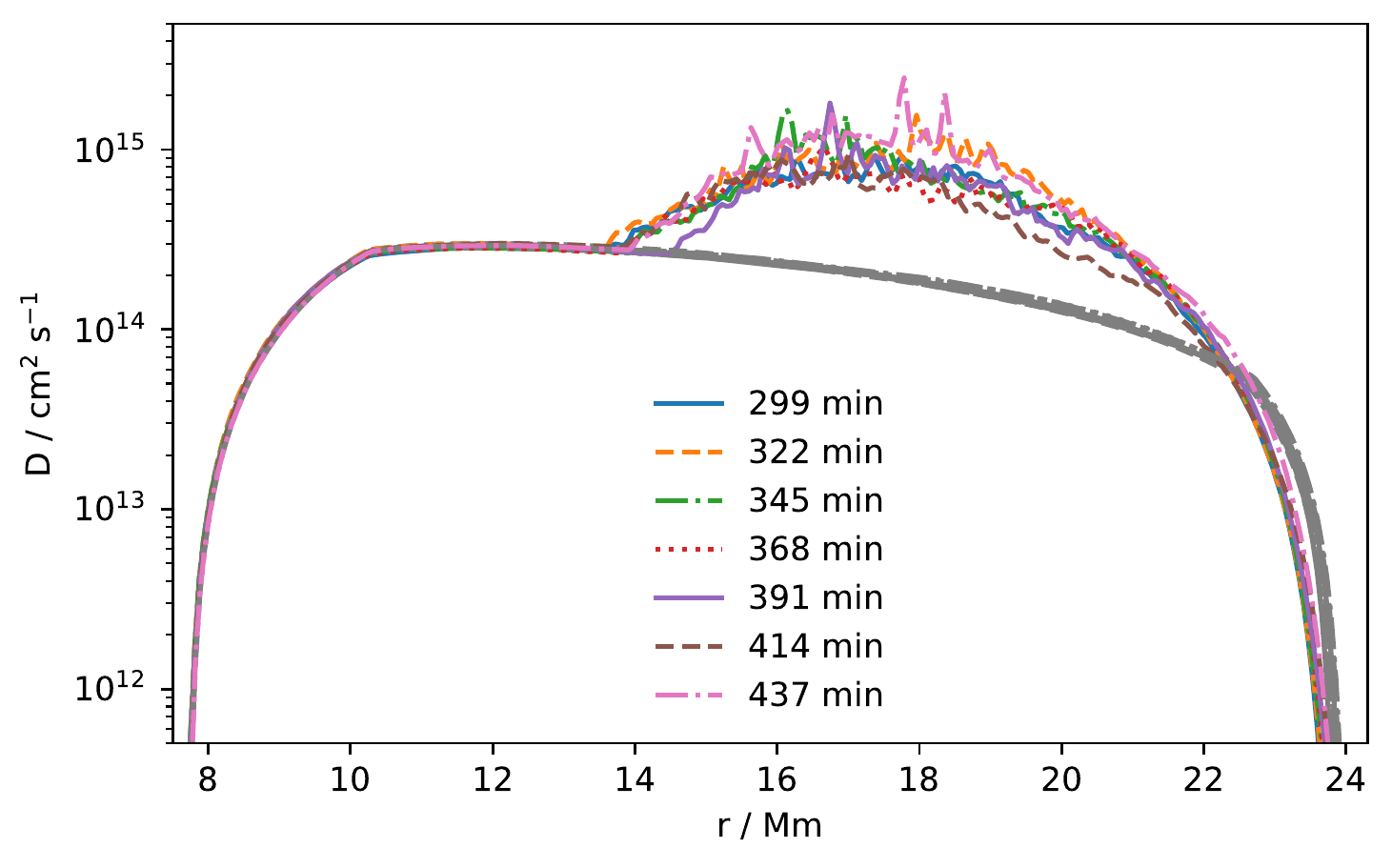}
  \caption{Top: $D_\mathrm{vr}$ according to \eq{dvr}, and
          $D_\mathrm{FV}$ according to the direct determination from the
          $X_\mathrm{H}$ profiles as well as the pressure scale
          height at $t = \unit{299}{min}$. Bottom: $D_\mathrm{vr}$ (grey lines) and the composite
          $D_\mathrm{FV+vr}$ (colour lines) according to the direct
          determination from the $X_\mathrm{H}$ at various times, merged
          with $D_\mathrm{vr}$ in the lower part of the convection zone (see
          text for details). }
  \label{fig:DvrDVF-compare}
\end{figure}

Despite taking the burning term into account in \eq{diff-inv-burn}, the profile
of the entrained material cannot be completely recovered, and thus the method
cannot provide a diffusion coefficient in the lower part of the convection zone
where all of the entrained material is burned (top panel \Fig{DvrDVF-compare}).
Therefore we also determine the diffusion coefficient from the
spherically averaged radial velocity $v_\mathrm{PPM}$ from the 3D
hydrodynamic simulation using the correction for the near-boundary
regions recommended by \citet{Jones:2017kc} based on O-shell
simulations
\begin{equation}
D_\mathrm{vr} = \frac{1}{3} v_\mathrm{PPM}\times \min \left(l,|r-r_0|\right),
\label{eq:dvr}
\end{equation}
where $v_\mathrm{PPM}$ is the rms of the radial velocity on spherical shells
from the 3D \ppmstar\ simulation of the He-shell convection, $l =
\alpha_{\mathrm{MLT}} H_P$ is the mixing length, and $|r-r_0|$ is the distance
to the He-shell or another convective boundary that is located at the radius $r_0$. For
$\alpha_{\mathrm{MLT}} = 1.6$, this prescription provides the best fit to the
diffusion coefficient derived from the hydrodynamic simulations of O-shell
convection simulations \citep[see Figure 21 in][]{Jones:2017kc}. However that
diffusion coefficient recommendation was focused on the behaviour of mixing near
and across the top convective boundary (CB), and no attempt was made to model
the discrepancy between the diffusion coefficient determined from the
spherically averaged 3D hydrodynamic abundance profile evolution and the
diffusion coefficient determined according to \eq{dvr}. This is clearly seen at
the middle of the convection zone, $r = \unit{5.5}{Mm}$, in Figure~21 of
\citet{Jones:2017kc}. In the RAWD case the burning of entrained H-rich material
takes place approximately in the middle of the convection zone where this
difference between $D_\mathrm{vr}$ and $D_\mathrm{FV}$ is largest ($\approx 0.4
\dots \unit{1.2}{dex}$, \Fig{DvrDVF-compare}).

Where we are able to determine $D_\mathrm{FV}$ it is the more accurate diffusion
coefficient that describes the underlying evolution of the spherically averaged
3D hydro profiles of these simulations better, for obvious reasons. In order to be able to use this
more accurate diffusion coefficient we combine it in the lower half with the
values of $D_\mathrm{vr}$ according to \eq{dvr}. We determine the radius
$r_\mathrm{Dmax}$ where $D_\mathrm{FV}$ has its maximum. Then
\begin{equation*}
  D_\mathrm{FV+vr}= \begin{cases}
  D_\mathrm{FV}, & \text{if } r > r_\mathrm{Dmax} , \\
  max(D_\mathrm{FV},D_\mathrm{vr}), & \text{otherwise}.
\end{cases}
\end{equation*}

\subsection{Advective mixing model}
\label{sec:advectmodel}

\begin{figure}
  \includegraphics[width=0.5\columnwidth]{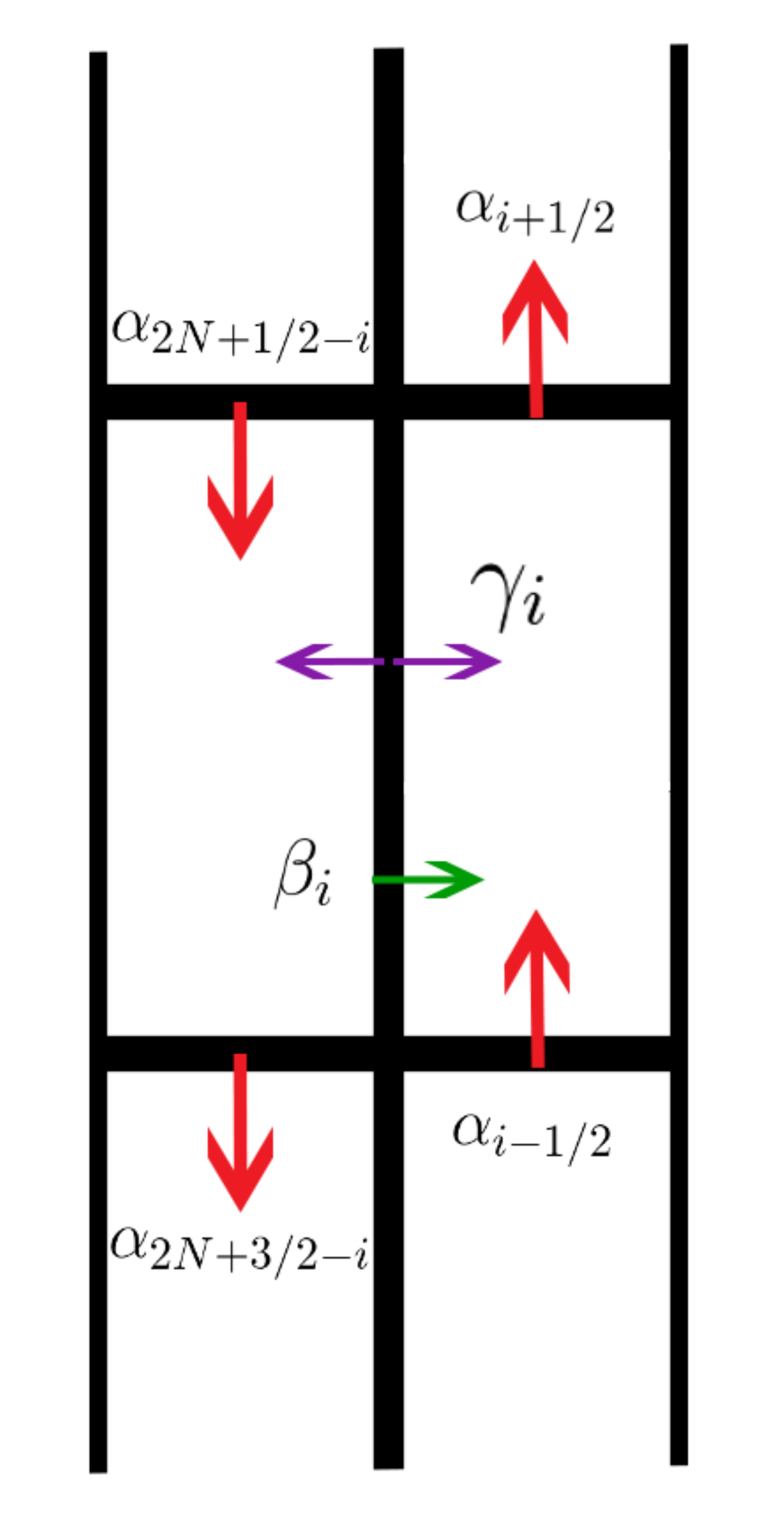}
  \centering
  \caption{An illustration of the two-stream model including all of the fluxes
          for a cell within the convection zone; $\alpha$ the radial mass flux
          (\Section{radialMix}), $\beta$ the enforced horizontal mass flux and $\gamma$
          the additional horizontal mass flux (\Section{horizontalMix}). The index $i$
          refers to the upstream cell which has an adjacent downstream cell $2N + 1 - i$.}
  \label{fig:elevator}

\end{figure}

The 1D advection is formulated using a two-stream approach in which
one stream transports species radially upwards, while the other stream
transports them radially
downwards \citep[][\Fig{elevator}]{cannon1993,Henkel2017}. Our
adaptation of this advective two-stream (ATS) model to post-processing
of 3D hydrodynamic simulations constitutes a reduced-dimensionality
1.5D advection model.

By default, the two streams are taken to be equal in their surface
areas which is justified by the analysis of the 3D flow properties
(see \Section{flow}) and the discussion in
\Section{radialMix}. The model can have horizontal mixing between up- and
downstream cells that are adjacent to each other.

There are $N$ cells per stream. The index $i$ refers to the spatial index of the
$2N$ cells and is used where the equations are agnostic to whether the cell is
within an up- or downstream. For computational and numerical simplicity, the
indexing is done such that the downstream is inverted and \textit{stitched} onto
the top of the upstream. The indexing starts at the bottom of the upstream at
$i=1$ and so cell $i=N+1$ is at the top of the downstream. The index $k$ refers
to species.

The discretization distinguishes variables defined on the cell boundaries which
have spatial half-integer indices, $i+1/2$, while those that represent cell
averages have spatial integer indices $i$.

\subsubsection{Discretized equations}
\label{sec:discretized}

To formulate the equations we start off with the conservation of mass equation
and then apply the divergence theorem to it to yield
\begin{equation}
  \label{eq:masscons}
  \partial_{t} \int_{V} \rho \mathrm{d}V = - \oint_{\partial V} \rho \vec{v} \cdot \mathrm{d}\vec{a}
\end{equation}
The term on the right hand side is interpreted as a sum of mass fluxes through
the boundary of the given volume, $V$. The term on the left hand side is the rate of
change of the mass contained within that volume, $V$. This advective model conserves
the total mass of every cell which requires that the sum of all mass fluxes at
every cell is zero. Recasting \eq{masscons} into the partial densities of every
species, integrating both sides and applying the constraint that the mass of
every cell is constant at all times leads to
\begin{equation}
  \label{eq:fluxes0}
  \partial_{t} \left( \sum_{k} m_{k,i} \right) = - (\sum_{k}\sum_{j=1}^{5} F_{k,i}^{j}) = 0
\end{equation}
where the index $j$ refers to a specific species mass flux on the
surface of cell $i$. The mass, $m_{k,i}$, of a given species within a
cell can change however the total mass within that cell cannot. This
is implicitly satisfied with \eq{fluxes0}. Therefore these equations
only transport the mass of a given species throughout the convection
zone. \Eq{fluxes0} includes all fluxes in the advective model and it
is integrated explicitly.

\subsubsection{Radial mixing and boundary conditions}
\label{sec:radialMix}

Within the convection zone, the bulk transport of species is through the radial
direction. The radial mass flux, $\alpha$, on each cell's boundary is given by
\begin{equation}
  \label{eq:alpha}
  \alpha_{i+1/2} = 2 \pi r^{2}_{i+1/2} \rho_{i+1/2} v_{i+1/2}
\end{equation}
where $r$, $\rho$, $v$ are the radius, density and velocity defined on
a cells boundary. The velocity is a positive-definite quantity in this
model; the species mass fluxes will explicitly carry the appropriate
sign for transport. In order for this model to remain consistent with
the nearly-hydrostatic equilibrium of the underlying 3D stellar
hydrodynamic simulation, the net mass flux at every radius must be
zero to ensure that there is no net mass transport in the radial
direction. This means that
\begin{equation}
  \label{eq:fluxlayer}
  \alpha_{i+1/2} = \alpha_{2N+1/2-i}
\end{equation}

In \eq{alpha}, the radial mass flux depends on a surface area, taken
to be $2 \pi r^{2}$, to which the underlying transport of mass, $\rho
v$, is advected through. In principle, the surface area can be
different between the up- and downstream cells and still ensure that
the net mass flux is zero so long as the product $\rho v$ with the
respective surface areas are constant at every radius.  With the low
\mach~number flows in the RAWD simulations, the largest density
perturbations are at the percent level as seen in \Fig{mollweide}. The
density can therefore be approximated to be constant at every radius
so that $\rho_{i+1/2} = \rho_{2N+1/2-i}$. The spherical average of the
radial velocity at every radius is approximately zero
(\Fig{mollweide}). The magnitude of the velocity at the surface of the
up- and downstream cells at each radius can then be approximated as
being the same, $v_{i+1/2} = v_{2N+1/2-i}$. With the density and
velocities being equal in both streams at a given radius, the surface
areas of each stream must also be equal to ensure that there is no net
radial mass transport. Another consequence of these approximations is
that the mass of the cells in both streams at a given radius are also
equal, $\delta m_{i} = \delta m_{2N+1-i}$.

The radial species mass fluxes for cell $i$ are
\begin{subequations}
\label{eq:Rfluxes}
\begin{align}
  F_{k,i}^{O} &= \alpha_{i+1/2} X_{k,i+1/2} \\
  F_{k,i}^{I} &= -\alpha_{i-1/2} X_{k,i-1/2}
\end{align}
\end{subequations}
where the superscript $O$ and $I$ refer to the outflow and inflow of mass at
cell $i$, respectively. These constitute two of the fluxes from \eq{fluxes0}.
The mass fractions, which are defined in the center of a cell, $X_{k,i}$, are
considered the average within that cell. To
achieve second order accuracy in the solutions of these equations, the mass
fraction on a boundary, $X_{k,i+1/2}$, is determined through a linear
interpolation using the neighbouring cell averages. There are
two estimates for the interpolated state,
\begin{subequations}
\begin{align*}
  X_{k,i+1/2} &= X_{k,i+1} - \frac{1}{2}\frac{\partial X_{k,i+1}}{\partial m_{i+1}}\delta m_{i+1} \\
  X_{k,i+1/2} &= X_{k,i} + \frac{1}{2}\frac{\partial X_{k,i}}{\partial m_{i}}\delta m_{i} 
\end{align*}
\end{subequations}
referring to the \textit{up-sided} and \textit{down-sided} estimates of the
interpolated mass fraction, respectively. To ensure that the numerical scheme is
stable, the interpolated state is chosen such that the discretization is
\textit{upwinding}. In practice, this just means that if the velocity is
upwards, the \textit{upwinded} interpolated state is the \textit{down-sided}
estimate because material is being advected upwards. To reduce oscillations in
the solutions the minmod limiter is used on the numerically estimated slope
\citep{LeVeque2002FiniteVM}.
\begin{equation*}
  \frac{\partial X_{k,i}}{\partial m_{i}} = \code{minmod}(\frac{X_{k,i} - X_{k,i-1}}{\delta m_{i}}, \frac{X_{k,i+1} - X_{k,i}}{\delta m_{i}})
\end{equation*}
where 
\begin{align*}
  \code{minmod}(a,b) = \left\{
  \begin{array}{lr}
      a & \text{if } |a| < |b| \text{ and } a \cdot b > 0 \\
      b & \text{if } |b| < |a| \text{ and } a \cdot b > 0 \\
      0 & \text{otherwise}
  \end{array}
  \right.
\end{align*}

At the inner and outer boundaries of the convection zone the velocities are
zero. Physically, this condition is enforcing that all of the radial flow is
turning over at the boundary such that there is only a horizontal flow. Of
course, fluid can and does over turn at some distance before the
boundary~\citep{Jones:2017kc} which are written as sources of horizontal mixing
in this model (\Section{horizontalMix}). The constraints on the radial mass flux
coefficients to enforce a horizontal flow at the boundaries are
\begin{subequations}
\begin{align*}
  &\alpha_{N-1/2} = \alpha_{N+1/2} = \alpha_{N+3/2} \\
  &\alpha_{2N-1/2} = \left( \alpha_{2N+1/2} \equiv \alpha_{1/2} \right) = \alpha_{3/2}
\end{align*}
\end{subequations}
These coefficients cause, at the uppermost cell in the upstream, all of the mass
that enters that cell to flow directly into the uppermost cell in the
downstream. This is a horizontal flow. These boundary conditions are essentially
periodic boundary conditions and from the numerical and computational
perspective, it is convenient to have the two streams attached to each other to
form a ring. By applying \eq{masscons} to a convection zone that has zero
velocity on its boundaries, there is no mass entering or leaving the convection
zone and so it remains constant. For this reason, the model follows the
Lagrangian coordinates of the \ppmstar{} initialized convection zone as it
expands in Eulerian coordinates (\Section{convbound}).

\subsubsection{Horizontal mixing}
\label{sec:horizontalMix}

With only the radial mass fluxes given by \eq{Rfluxes} contributing to
\eq{fluxes0} thus far, the only way in which the fluxes at the upper and lower
boundary of any cell sum to zero is if the product of $2 \pi r^{2} \rho v$ is
constant for all radii. This is not true for the convection zone in the RAWD
simulations nor in general. Rather, the rms radial velocity
profile from the 3D simulations in \Fig{rprofs} shows a pronounced peak about
1/3$^\mathrm{rd}$ from the bottom of the convection zone and falls off well
inside the CBs where the flow turns around in a broad sweep corresponding to
the low \mach~numbers.

Using \Fig{elevator} as a reference, if the upstream has $\alpha_{i+1/2} >
\alpha_{i-1/2}$, then over a time step the mass within that cell $i$ will
decrease. Simultaneously the downstream will be increasing its mass by the same
amount due to the radial mass fluxes being equal in the two streams at every
radius (\eq{fluxlayer}). To conserve the mass in each of the cells, there is an
enforced horizontal mass flux of
\begin{equation}
  \label{eq:beta}
    \beta_{i} = \alpha_{i+1/2} - \alpha_{i-1/2}
\end{equation}
from one stream to the other. The sign of this coefficient $\beta$ determines
whether mass is transferred from the downstream to the upstream or vice versa.
The horizontal species mass flux at the upstream cell with index $i$ is
\begin{align*}
  F^{\beta}_{k,i} = \left\{
  \begin{array}{lr}
      - | \beta_{i} | X_{k,2N+1-i} & \text{if } \beta_{i} > 0 \\
       | \beta_{i} | X_{k,i} & \text{if } \beta_{i} < 0
  \end{array}
  \right.
\end{align*}

One could consider a situation where neighboring cells at the same radius
conserve mass in the shell but allow for the mass in each cell to be variable so
that the $\beta$ coefficient is not exactly as written in \eq{beta}. However,
since the density fluctuations are at the percent level as shown in
\Fig{mollweide} they are approximated as being the same between the two streams at every radius.
Therefore the mass within each cell should not change due to horizontal mixing.
Combining this fact with there being no net radial transport of mass, the mass
within every cell is constant at all times.

With the $\beta$ and $\alpha$ mass fluxes, the total mass flux at every cell is
zero. There can be additional horizontal mass fluxes so long as the sum of them
is equal to zero. This additional horizontal mass flux is given the symbol
$\gamma$ and it is unique within each shell, i.e $\gamma_{i} =
\gamma_{2N+1-i}$, to ensure that the total mass flux at every cell is still
zero. However, the additional horizontal species mass flux is not zero and has
the form
\begin{equation}
\label{eq:gammaflux}
  F^{\gamma}_{k,i} = \gamma_{i} \left( X_{k,i} - X_{k,2N+1-i} \right)
\end{equation}
for each cell.

While mass is being transported radially in the up- and downstreams, mass is
also being exchanged between them through a horizontal mass flux. The mixing of
the mass of a given species $k$ between cell $i$ and its adjacent cell $2N+1-i$
implied by \eq{gammaflux} has the effect of homogenizing the species between the
two streams at every layer, depending on the value of $\gamma$. There are two
competing timescales that determine how efficient the mixing between the up- and
downstreams are. These are the radial advection timescale,
\begin{equation}
  \label{eq:radialT}
  \delta t_{r,i} = \frac{|\delta r_{i}|}{v_{r,i}}
\end{equation}
and the horizontal time scale, $\delta t_{h,i}$, for a given cell. Associated
with these timescales are mass fluxes. The radial mass flux is $\alpha$, while
the horizontal mass flux is $\gamma$. With these two competing timescales, if
one is shorter than the other then it is implied that the shorter timescale has
a larger mass flux than the longer timescale. This can be written as
\begin{equation}
  \label{eq:timescales}
  \gamma_{i} = \alpha_{i} \frac{\delta t_{r,i}}{\delta t_{h,i}}
\end{equation}

The mental picture of a convection zone being composed of one upstream and one
downstream that splits the convection zone evenly into two hemispheres is very
unrealistic (\Fig{mollweide}). Instead, this model should be thought of as the
upstream being the superposition of all radially upward flows and similarly for
the downstream. From this idea, if the upward flows are small in their surface
area at a given radius, they can more easily mix with the adjacent downward
flows. To decompose the velocity into upward and downward flows at a particular
length scale on a sphere, the power spectrum (spherical harmonics) of the radial
velocity is taken. To compute the spherical harmonics the python package
\code{pyshtools} is used for data sampled on a sphere. The maximum mode, $\ell$,
that the velocity field is decomposed into is chosen such that the smallest
wavelength, $\lambda = 2 \delta r$ where $\lambda = 2 \pi r / \sqrt{\ell(\ell +
1)}$, that can be sampled from the \textit{briquette} data (see \Fig{mollweide}
and
\Section{RAWD_Hydro}) with a grid spacing of $\delta r$ is computed.

For a given mode, $\ell$, there is an associated length scale, or wavelength. A
particle at the center of a stream must move, tangentially to the surface of the
sphere, half of this wavelength to be in the center of the adjacent stream. With
an average tangential velocity, the particle will take $\delta t_{h,i,\ell} =
\pi r_{i} / \left( \langle|v_{\perp,i}|\rangle\sqrt{\ell(\ell+1)} \right)$
amount of time to mix, for a given mode $\ell$. The flow does not develop with
only one very dominant mode but a spectrum of modes (\Fig{power-spectrum}) and
so the horizontal timescale is weighted by its power, $S_{cc}(\ell,r) =
\sum_{m=-\ell}^{\ell}|c_{\ell,m}(r)|^{2}$. Therefore, the horizontal timescale
is
\begin{equation}
  \label{eq:horizontalT}
  \delta t_{h,i} = \sum_{\ell} \frac{S_{cc}(\ell,r_{i}) \pi r_{i}}{\left( \sum\limits_{\ell} S_{cc}(\ell,r_{i}) \right) \langle |v_{\perp,i}| \rangle\sqrt{\ell(\ell+1)}}
\end{equation}

\subsubsection{Entrainment}
\label{sec:entrain}

A convection zone can entrain, or ingest, material due to hydrodynamic
instabilities at the boundaries. In the RAWD simulations, the \cldfluid{} 
is entrained into the convection zone through a downstream at the convective
boundaries, as shown in \Fig{mollweide}, bringing H-rich material into the
convection zone. With the constraint on the advective model that the mass within
any cell cannot change, the mass of the \cldfluid{} that is entrained into
a cell must also be removed. Algorithmically, the entrainment process follows
what was discussed in detail in Appendix A of \citet{Denissenkov:19}. The H-rich
material is ingested into the downstream into a region of mass, $\Delta M$, that
is equal to the mass of the top cell of the downstream. This ensures that all of
the mass that is entrained is done so directly through the convective boundary
surface, thus incorporating the horizontal and radial mixing timescales to
determine how fast the composition changes are transported downwards. Due to
$\Delta M$ being only one cell, the change in the mass fractions is applied as a
step increase.

To test how well entrainment and mixing is modeled we have switched off
all reactions and compared the total amount of H ingested into the top cell with
the amount of H distributed over the other cells by mixing. The relative
difference between these amounts in the advective model remains within $2-4\%$
even after 30334 time steps (\unit{744}{min} of integration time). In the
diffusive model the entrained material is distributed evenly over 18 cells (out
of a total number of 256 cells) using the step increase method from Appendix A
of \citet{Denissenkov:19}. Over the same number of time steps as the advective
model we find that the relative differences between the total entrained H and
the mixed H in the diffusive model to be of the same order, $2-4\%$.

\subsubsection{The Courant Condition}
\label{sec:courant}

With the fact that the method is explicit, the time steps are limited by the
Courant condition. With the cells being discretized in mass, the Courant
condition can be plainly stated as requiring that the total amount of mass
advected out of any given cell in a single time step cannot be larger than the
mass within that cell. The Courant number in any cell is given by
\begin{equation}
  \label{eq:courant}
  C_{i} = \frac{\delta t \left(\alpha_{i+1/2} + \gamma_{i} + |\beta_{i}| \right)}{\delta m_{i}}
\end{equation}
and the Courant condition states that for all $i$, $C_{i} \leq 1$. Any advective
post-processing simulations have the condition that the maximum Courant number
across all cells is $0.5$. This determines the time steps that any
post-processing model takes (see \Section{modelConstraints}).

\subsection{\mppnp{} post-processing simulations}
\label{sec:mppnpSims}

The 1D multi-zone NuGrid code, \mppnp{}
\citep{Pignatari:2016er,RitterSE}, is used for the 1D post-processing
models of the \ppmstar{} simulations. Either the diffusive or
advective mixing routine, as described in
\Sections{Dmethod}{advectmodel} respectively, is used for a given
\mppnp{} simulation which are tabulated in \Tab{mppnp_runs}. The first
set of runs has been performed by including only the
$\nuclei{12}{C}(\proton,\gamma)\nuclei{13}{N}$ reaction rate and
neglecting its electron screening, just as in the \ppmstar{} runs, in
order to have a direct comparison between 1D and 3D. The second set of
runs has been performed with a full network of $\sim$1000 species
suitable for \ipr\ simulations in which we use the same nuclear data
and electron screening factors as in \citet{Denissenkov:19}. Both the
advective and diffusive mixing post-processing models, for a given 3D
\ppmstar{} run, use the same spherically averaged radial $\rho$, $T$
profiles and time steps corresponding to the 3D dump times. The
details of the formulation of the models are discussed
in \Section{modelConstraints}.

\begin{table}
  \centering
  \caption{Summary of the 1D \mppnp{} post-processing simulations.}
	\label{tab:mppnp_runs}

  \resizebox{\columnwidth}{!}{%
    
  \begin{threeparttable}
    \begin{tabular}{llll} 
      \toprule
      \mppnp{} Run ID & A/D$^i$            & Hydro Run ID & Network$^{ii}$ \\
      \midrule
      mp1             & $D_\mathrm{FV+vr}$ & N16          & $\nuclei{12}{C}(\proton,\gamma)\nuclei{13}{N}$\\
      mp2             & ATS               & N16          & $\nuclei{12}{C}(\proton,\gamma)\nuclei{13}{N}$\\ 
      mp3             & $D_\mathrm{vr}$    & N17          & $\nuclei{12}{C}(\proton,\gamma)\nuclei{13}{N}$\\
      mp4             & ATS               & N17          & $\nuclei{12}{C}(\proton,\gamma)\nuclei{13}{N}$\\ 
      mp5             & $D_\mathrm{FV+vr}$ & N16          & full\\ 
      mp6             & ATS               & N16          & full\\ 
      mp7             & $D_{\mathrm{vr}}$  & N17          & full\\ 
      mp8             & ATS               & N17          & full\\ 
      mp9             & $D_{\mathrm{vr}}$  & N16          & $\nuclei{12}{C}(\proton,\gamma)\nuclei{13}{N}$\\
      mp10            & $D_{\mathrm{vr}}$  & N16          & full\\
      \bottomrule
    \end{tabular}
  \begin{tablenotes}
    \small
    \item $^i$ The mixing model used in the post-processing which is either
          diffusive ($D_{\mathrm{vr}}$ or $D_{\mathrm{FV+vr}}$, \Section{Dmethod}) or
          advective (ATS, \Section{advectmodel}); $^{ii}$ $\nuclei{12}{C}(\proton,\gamma)\nuclei{13}{N}$
          indicates post-processing with the same one-reaction burn that is implemented in
          the \ppmstar{} runs, full indicates post-processing with the complete network
          needed for \ipr{} modeling
  \end{tablenotes}
\end{threeparttable}
}
\end{table}


\section{Results}
\label{sec:results}

We first describe the hydrodynamic simulations of RAWDs in \Section{RAWD_Hydro}.
The details of the post-processing models are discussed in \Section{modelConstraints}.
Focusing on the only nuclear reaction included in the hydrodynamic simulation we
then describe and compare post-processing via the diffusive (\Section{Dc12pg})
and advective methods (\Section{Advc12pg}). In \Section{fulllnetworkresults} we
describe the post-processing results including a full network and comparison
with observations. 

\subsection{3D hydrodynamic simulations of RAWDs}
\label{sec:RAWD_Hydro}
\subsubsection{Flow properties}
\label{sec:flow}

\begin{figure}
  \includegraphics[width=\columnwidth]{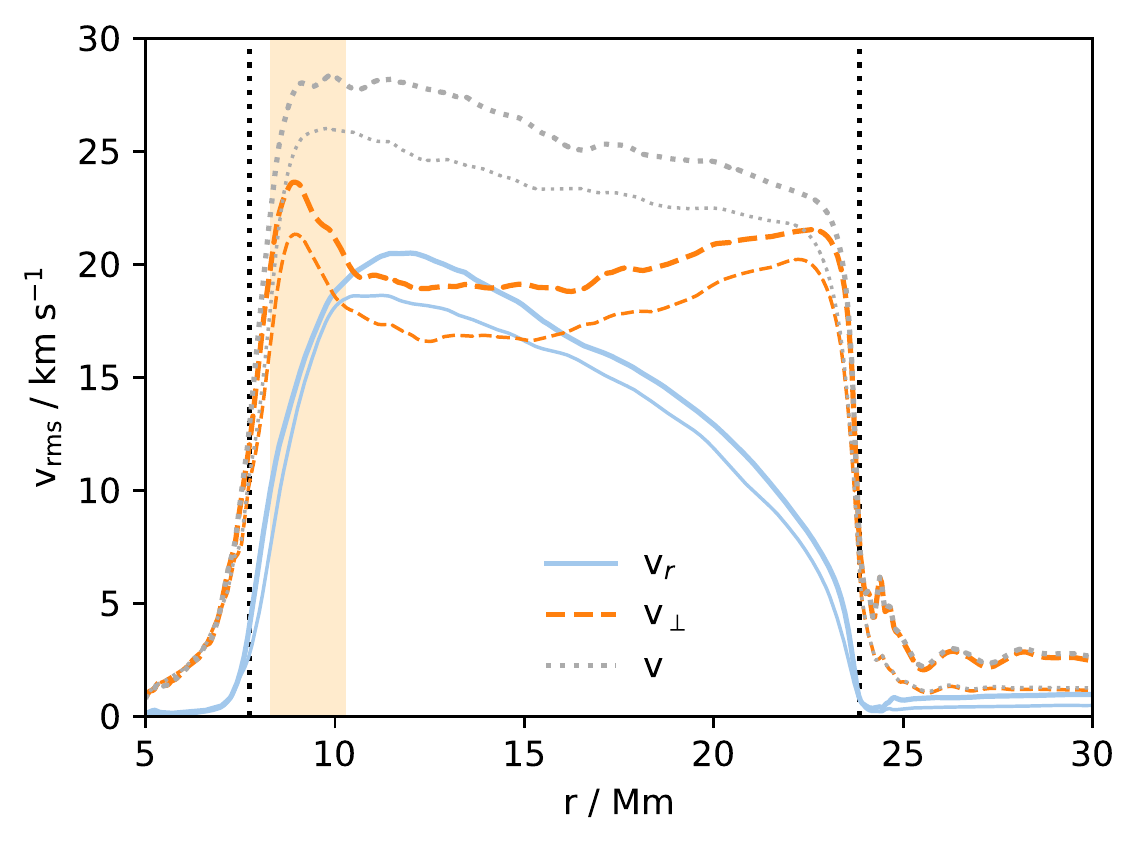}
  \centering
  \caption{The rms radial profiles of $v_{r}$, $v_{\perp}$ and
          $v$ at $t = \unit{299}{min}$. The thick lines are from run \code{N16}, while
          the thin lines are from run \code{N15}. The black dotted lines correspond to the
          convective boundaries that are used in \code{N16}'s post-processing models
          (\Sections{convbound}{modelConstraints}). The beige shaded region is where the
          volume heating is applied. When spatially integrated it corresponds to the
          helium luminosity, $L_{\mathrm{He}}$, stated in \Tab{ppmstarmodels}.}
  \label{fig:rprofs}

\end{figure}

As with the other \ppmstar{} simulations in
\citet{woodward15,Jones:2017kc,Andrassy:19}, the RAWD simulations do not
start with any initial perturbations in the velocity field but rather the
numerical representation of the spherically symmetric thermodynamic variables
develop instabilities. These are quickly over taken by the flow developing from
heat being injected (\Tab{ppmstarmodels}) in a thick spherical shell. This shell
is contained within the radii $\unit{8.28}{Mm}$ and $\unit{10.28}{Mm}$ and can
be seen as the beige shaded region in \Fig{rprofs}. After about 30 minutes this
transient state is completely lost in the simulations and the natural flow of
the convection zone has developed.

\begin{figure*}

  \begin{minipage}{.49\textwidth}
    \centering
    \includegraphics[width=\textwidth]{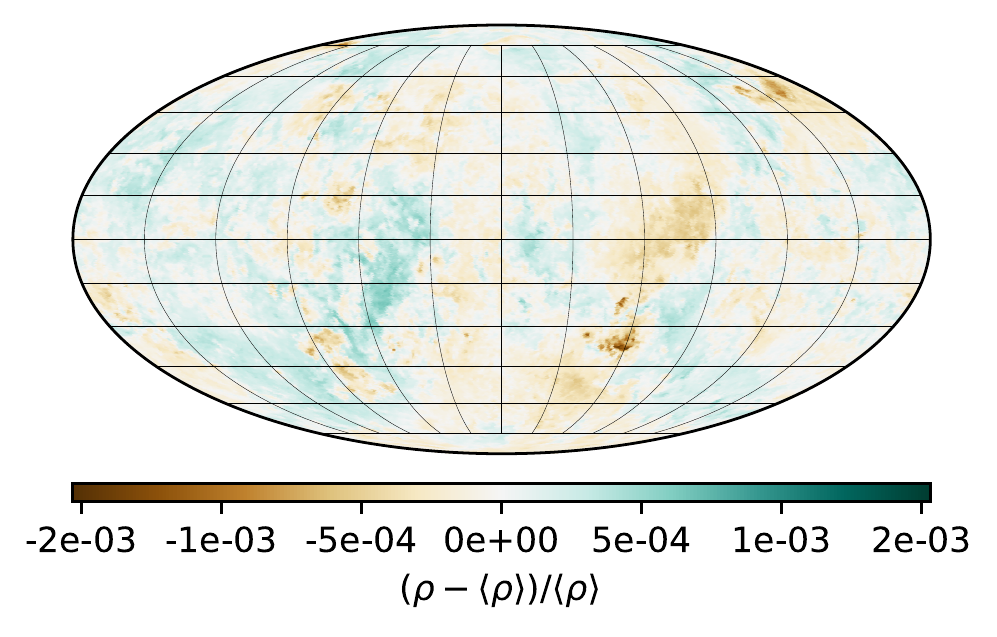}
  \end{minipage}
  \begin{minipage}{.49\textwidth}
    \centering
    \includegraphics[width=\textwidth]{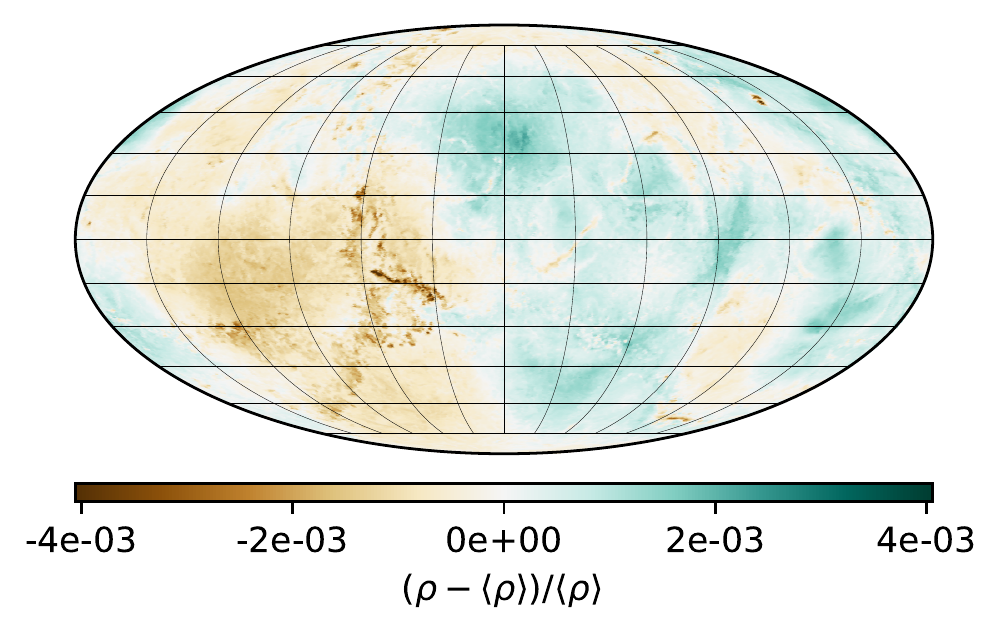}
  \end{minipage} \\

  \begin{minipage}{.49\textwidth}
    \centering
    \includegraphics[width=\textwidth]{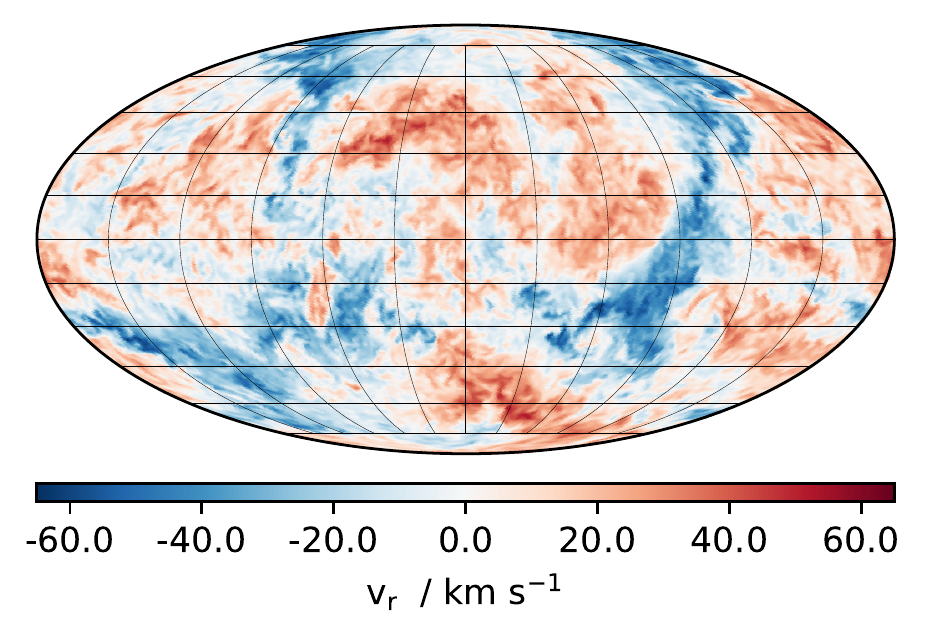}
  \end{minipage}
  \begin{minipage}{.49\textwidth}
    \centering
    \includegraphics[width=\textwidth]{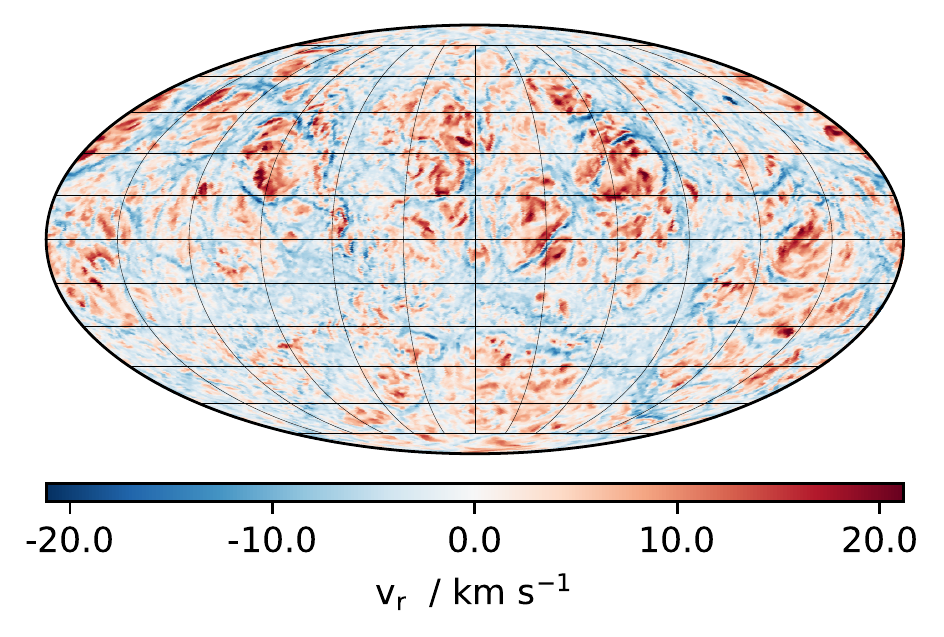}
  \end{minipage} \\
  
  \begin{minipage}{.49\textwidth}
    \centering
    \includegraphics[width=\textwidth]{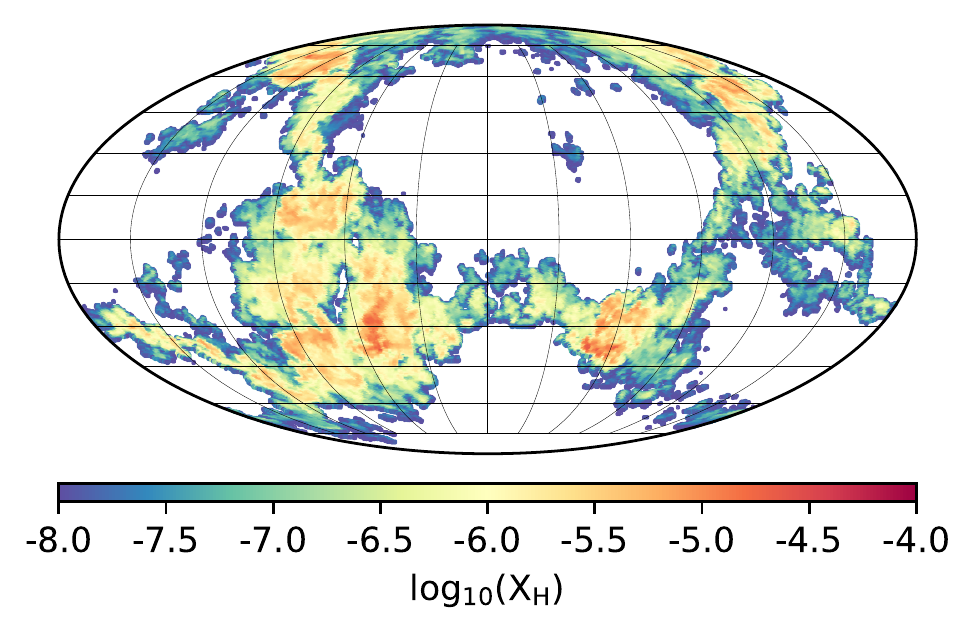}
  \end{minipage}
  \begin{minipage}{.49\textwidth}
    \centering
    \includegraphics[width=\textwidth]{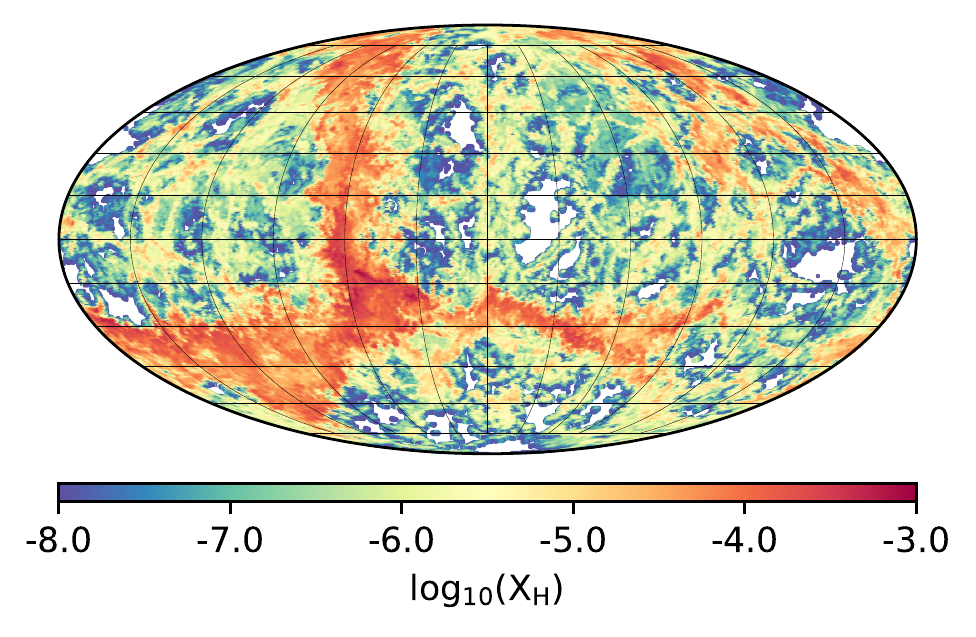}
  \end{minipage}

  \caption{In each row the left panel is a Mollweide plot of the quantity at a
          radius of $\unit{14.5}{Mm}$, which is within the H burning region
          (\Fig{burningrate}), while the right panel is at a radius of $\unit{23}{Mm}$
          which is near the top CB (\Tab{ppmstarmodels}). The first row plots the density
          perturbations from its mean, the second row plots the radial velocity, while the
          last row plots the mass fraction of H which is calculated from the FV that is at
          double the grid resolution. The Mollweide plots were made with the
          \textit{briquette} data which is downsampled by a factor of 4 (384$^{3}$) from
          the resolution of the underlying run, \code{N16} (1536$^{3}$), and are taken at
          $t = \unit{299}{min}$. The points on the shell to which the quantity is sampled
          on are distributed such that each point's surface area coverage is roughly equal
          to $4 \pi / N$. The number of points that are used roughly corresponds to having
          a single point per cubic cell.}
  \label{fig:mollweide}

\end{figure*}

The rms radial profiles of the radial and tangential velocities are shown in
\Fig{rprofs}. The radial velocity drops sharply near the CBs, while the
tangential velocity is dominant near the CBs. Directly from these profiles it
is clear that there will be significantly more horizontal mass transport near
the CBs (\eq{timescales}) than the middle of the convection zone. This
coincides with the flows being forced to turn over near the CBs. \code{N15}'s
spherically averaged velocities are smaller than \code{N16}'s and it is
significant when considering that the difference between the two runs is the
doubling of the spatial resolution in \code{N16}. A possible cause for this is
due to the higher entrainment rates in \code{N15} compared to \code{N16}, shown
in \Fig{entrain}. With the convective fluids doing work to bring the initially
stable \cldfluid{} into the convection zone, some of its kinetic energy is
lost resulting in the lower velocities.

\begin{figure*}

  \begin{minipage}{.49\textwidth}
    \centering
    \includegraphics[width=\textwidth]{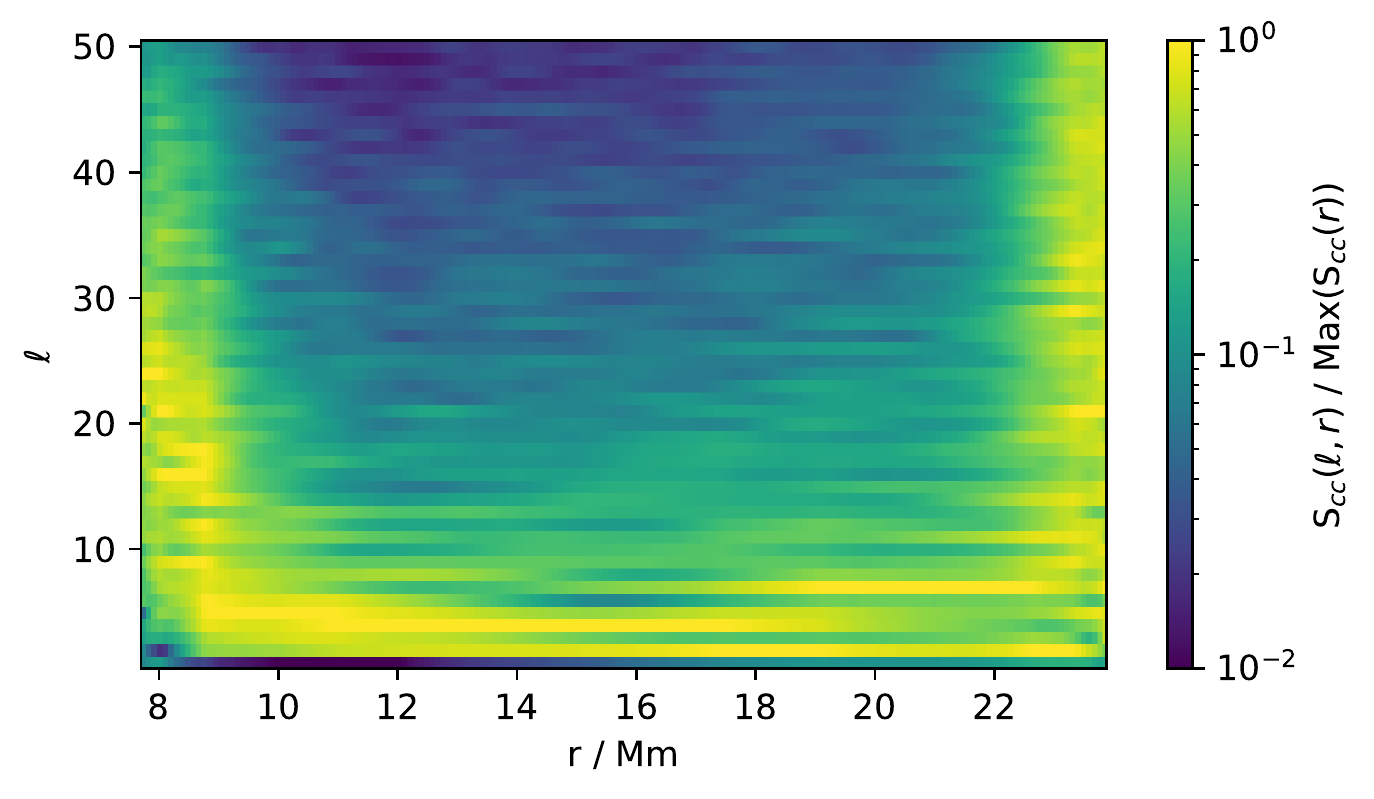}
  \end{minipage}
  \begin{minipage}{.49\textwidth}
    \centering
    \includegraphics[width=\textwidth]{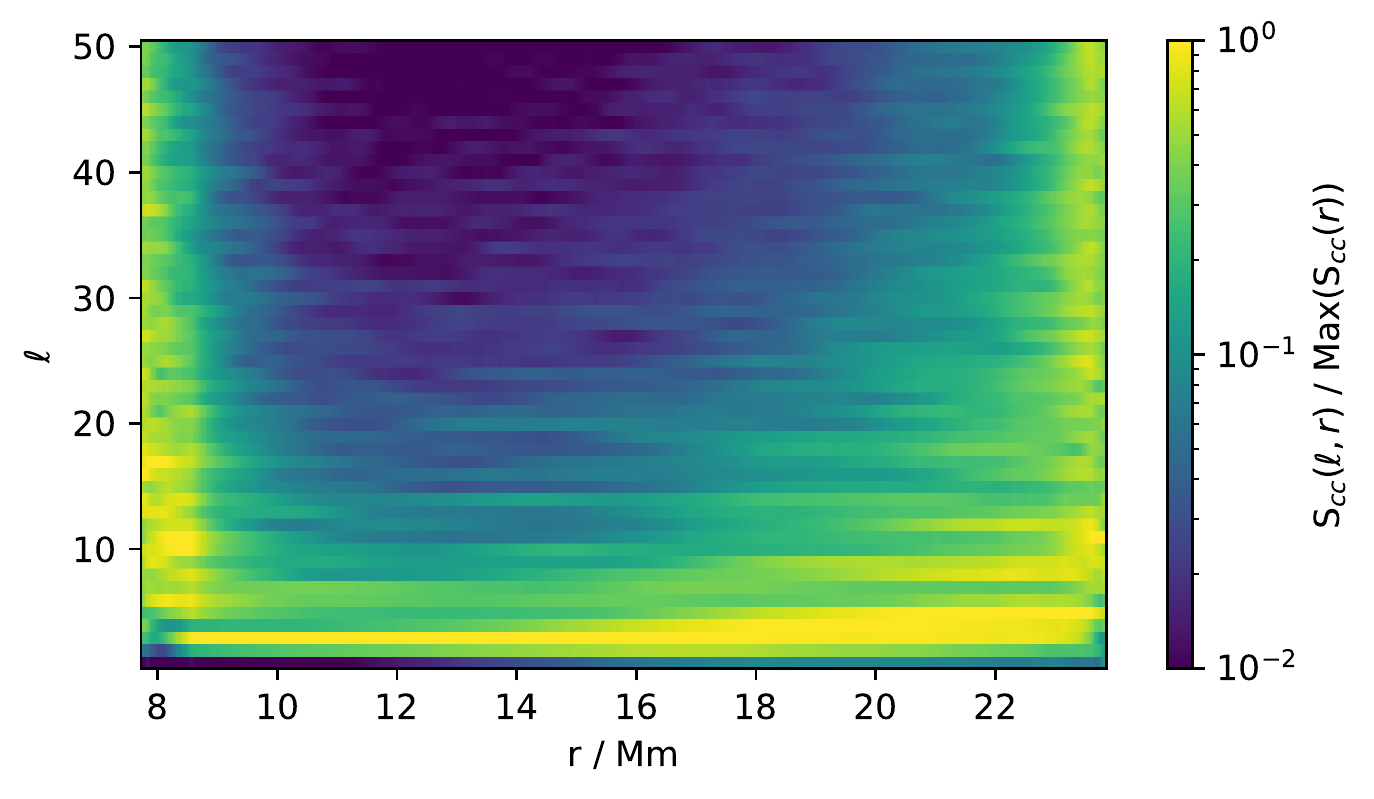}
  \end{minipage}
  \caption{The power spectrum, in terms of the spherical harmonic modes, $\ell$,
          of the radial velocity as a function of radius at $t = \unit{299}{min}$ within
          the convection zone. The left panel is from run \code{N15}, while the right panel
          is from run \code{N16}. The power in each radial bin is normalized by the
          maximum power within that radial bin.}
  \label{fig:power-spectrum}

\end{figure*}

\begin{figure}

  \includegraphics[width=\columnwidth]{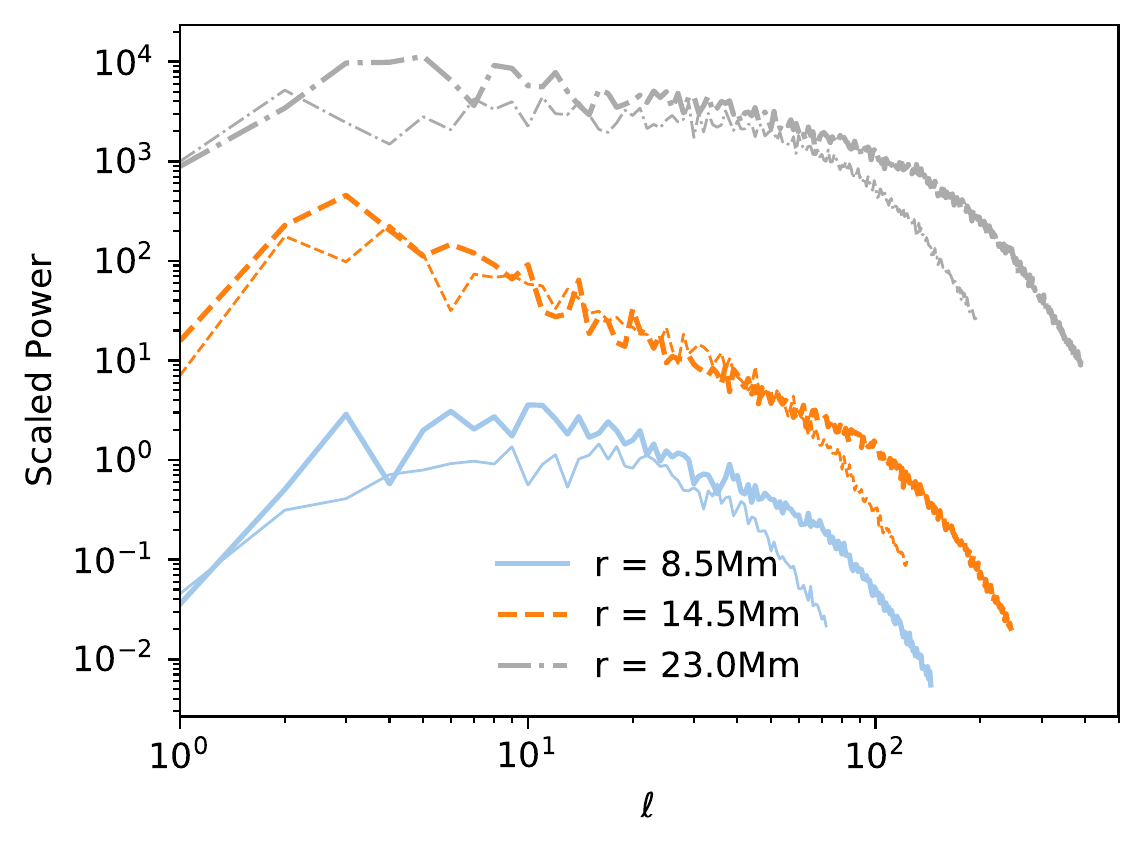}
  \centering
  \caption{The power spectrum of the radial velocity as a function of the spherical harmonic mode,
          $\ell$, at select radii at $t = \unit{299}{min}$. The
          thick lines are run \code{N16}, while the thin lines are from run \code{N15}. The
          spectrum is limited by the Nyquist sampling of the \textit{briquette} data
          resulting in more $\ell$ modes being resolved in the higher resolution run,
          \code{N16}. The power in each spectrum at a given radius is scaled by an
          arbitrary constant factor, which is the same for each hydro run, in
          order to show their $\ell$ dependence clearly.}
  \label{fig:lspectrum}

\end{figure}

The radial velocity field is shown at two radii in \Fig{mollweide}. These
Mollweide-projection plots are made using the \textit{briquette} data from a
\ppmstar{} simulation. This data set is downsampled in each spatial direction by
a factor of four from the original simulation through averaging the data in
4$^{3}$ cells. The radial velocity at $\unit{14.5}{Mm}$ is mostly dominated by
two modes, $\ell = 2$ and $3$, which can be seen visually in \Fig{mollweide} as
well as from its power spectrum in \FigTwo{power-spectrum}{lspectrum}. This is
consistent with the flow being dominated by the most unstable and largest
convective mode that can be in that spherical shell, $\ell \approx \pi r /
\Delta r \approx 3$~\citep{chandrabook1961}. At $\unit{23}{Mm}$, the velocity
field is not dominated by a few modes but the power is instead spread over many
modes of $\ell$. The large plumes that are advecting from the center of the
convection zone are broken up into smaller, incoherent streams that are swept
across by the large tangential velocities (\FigTwo{rprofs}{vt-mollweide}).

\begin{figure*}

  \begin{minipage}{.49\textwidth}
    \centering
    \includegraphics[width=\textwidth]{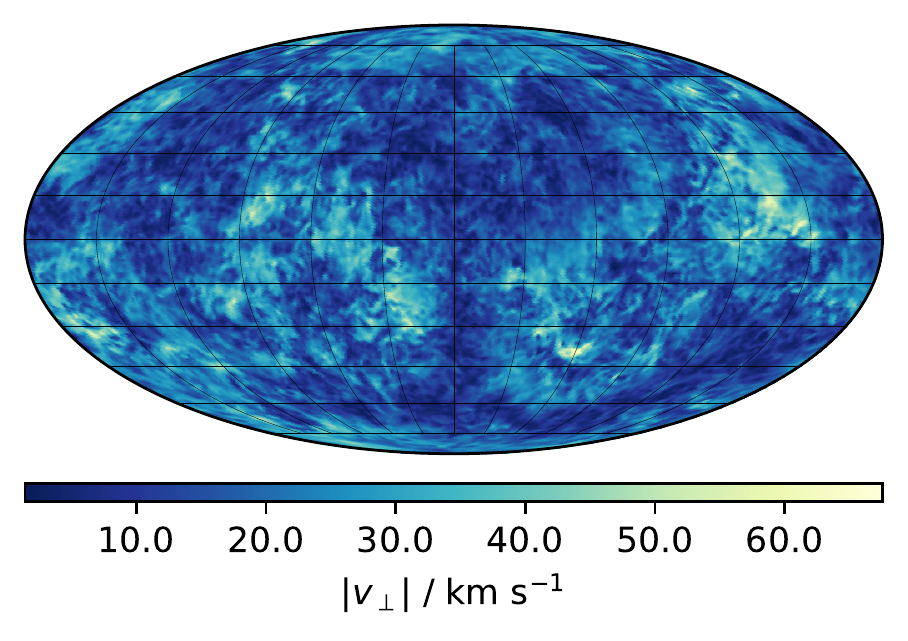}
  \end{minipage}
  \begin{minipage}{.49\textwidth}
    \centering
    \includegraphics[width=\textwidth]{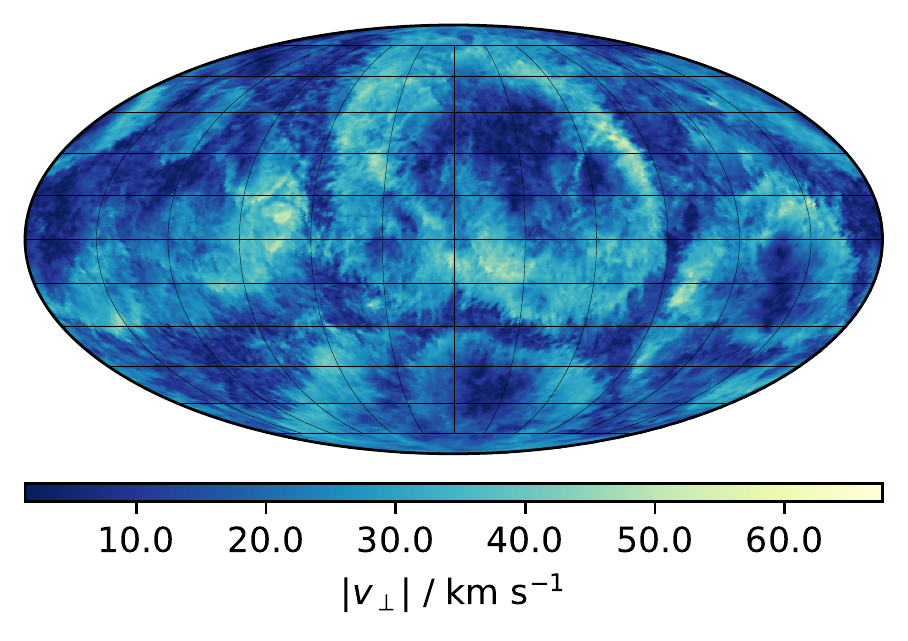}
  \end{minipage}
  \caption{The magnitude of the tangential velocity, $|v_{\perp}| =
          \sqrt{v_{\phi}^{2} + v_{\theta}^{2}}$, on spherical shells from \code{N16}.
          The left panel is at a radius of $\unit{14.5}{Mm}$, within the H burning region
          (\Fig{burningrate}), while the right panel is at a radius of $\unit{23}{Mm}$
          which is near the upper convective boundary (\Tab{ppmstarmodels}). These
          snapshots are taken at $t = \unit{299}{min}$.}
  \label{fig:vt-mollweide}

\end{figure*}

\subsubsection{The convective boundary and entrainment and burning of H}
\label{sec:convbound}

\begin{figure}

  \includegraphics[width=\columnwidth]{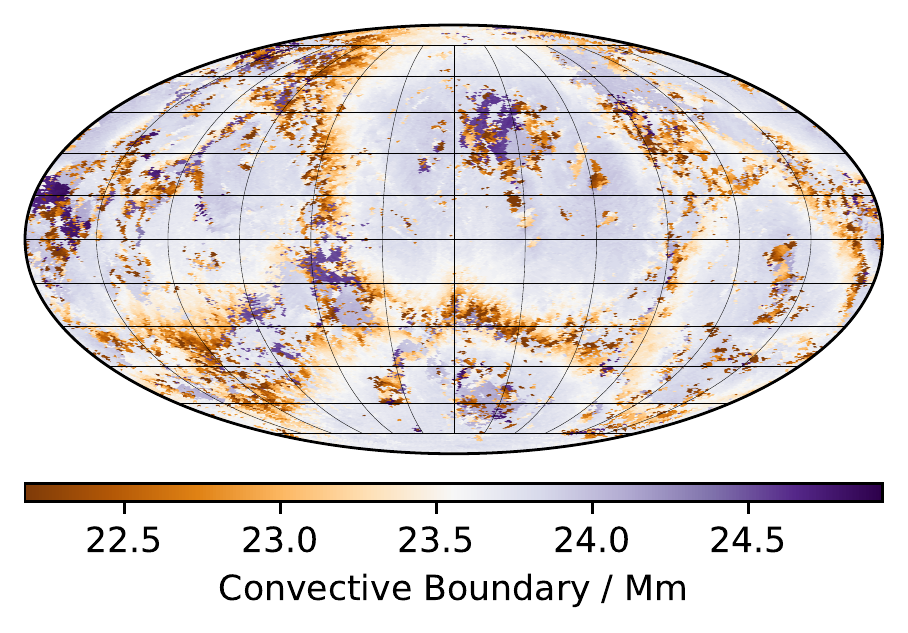}
  \centering
  \caption{Location of the upper convective boundary of \code{N16} according to
          the minimum of $\partial v_{\perp} / \partial r$ at $t = \unit{299}{min}$ from
          the \textit{briquette} data. Taking the spherical average, the upper convective
          boundary is at $\unit{23.56}{Mm}$ with $\sigma_{r_{\mathrm{b}},
            v_{\perp}} = \unit{0.46}{Mm}$. This is consistent with the other CB
          determination which uses the mass coordinates of the initial Schwarzschild
          boundary to yield an upper CB at $\unit{23.84}{Mm}$.}
  \label{fig:vtbound}

\end{figure}

The CB in 1D stellar evolution models can be determined by the Schwarzschild
criterion. If the fluid is unstable to convection and is fully mixed within the
convection zone, it is adiabatically stratified such that $dS/dr = 0$, or in the
case of an ideal gas equation state, which is used for the simulations done in
this paper, it can be expressed as $dA/dr = 0$ where $A = P / \rho^{\gamma}$
with $\gamma = 5/3$. The \ppmstar{} simulations are initialized with a
convection zone defined by these properties. The CB can be determined throughout
the entirety of the simulation by assuming that the convection zone is only
expanding due to heat and so it can be determined directly by using the
Lagrangian coordinates of the initial Schwarzschild boundary. This initial
Schwarzschild boundary is determined numerically with the condition that fluid
with $dA/dr > 0$ is stably stratified. In previous works that used the \ppmstar{}
code \citep{Jones:2017kc, Andrassy:19,Denissenkov:19} the CB during the
simulation was determined using the minimum of $\partial v_{\perp} / \partial r$
with the spherically averaged profiles of $|v_{\perp}|$ or with the
\textit{bucket} data of $v_{\perp}$ (see Section~3.3 in \citet{Jones:2017kc}).
The Schwarzschild criterion does not adequately describe the stability of the
fluid near the boundaries in these simulations as even in areas where the
entropy gradient is weakly positive, the fluid is flowing with moderate
velocities \citep{Jones:2017kc}. The gradient condition expresses the 3D nature
of the CB as when the fluid advects near the stiff boundary, the fluid is forced
to turn over. This can be viewed as small perturbations from a spherically
symmetric boundary as shown in \Fig{vtbound}. The thickness of this CB as
described by 1$\sigma$ spatial fluctuations~\citep[Fig.~17 in][]{Jones:2017kc}
is approximately $\unit{0.5}{Mm}$, which is smaller than the pressure scale
height at that boundary, $H_{P} = \unit{1.0}{Mm}$. The CB as determined by the
minimum of $\partial v_{\perp} / \partial r$ as well as the CB defined by the
Lagrangian coordinates of the initial Schwarzschild boundary for all runs at $t
= \unit{299}{min}$ are tabulated in \Tab{ppmstarmodels}. A discussion on the
appropriate boundary for the \mppnp{} post-processing models is
within \Section{modelConstraints}.

\begin{figure}

    \includegraphics[width=\columnwidth]{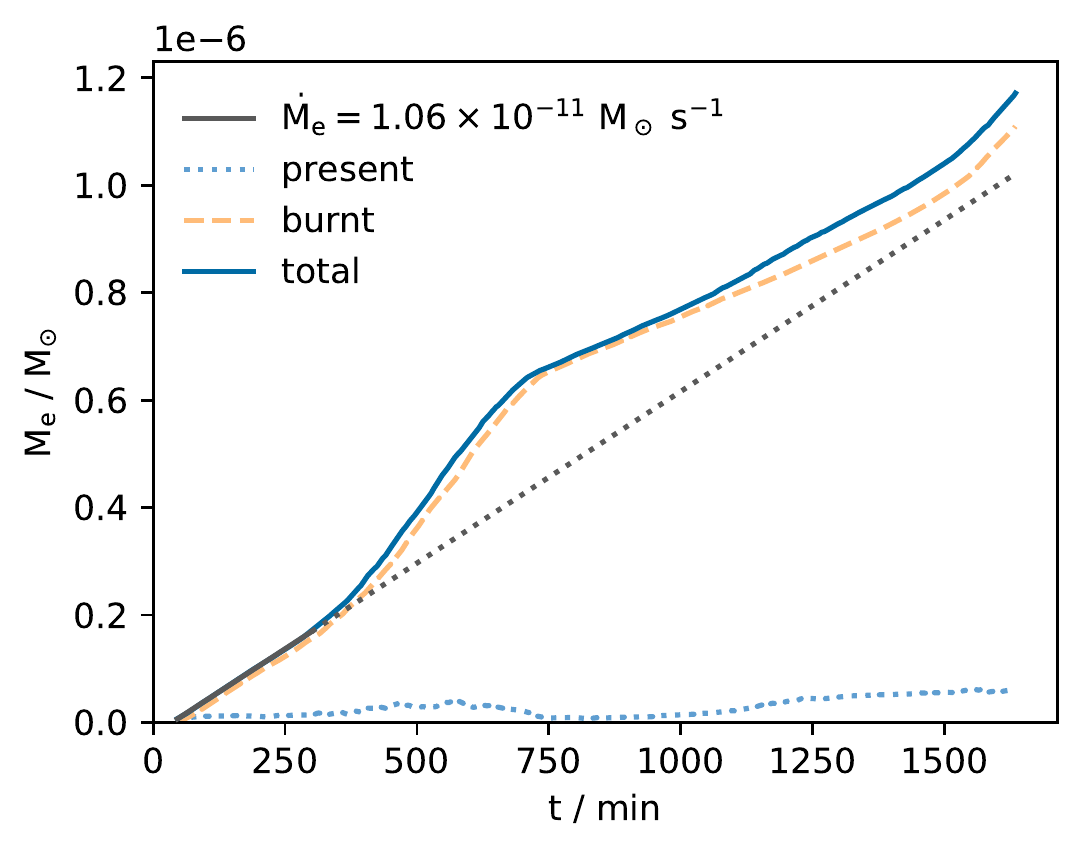}
    \includegraphics[width=\columnwidth]{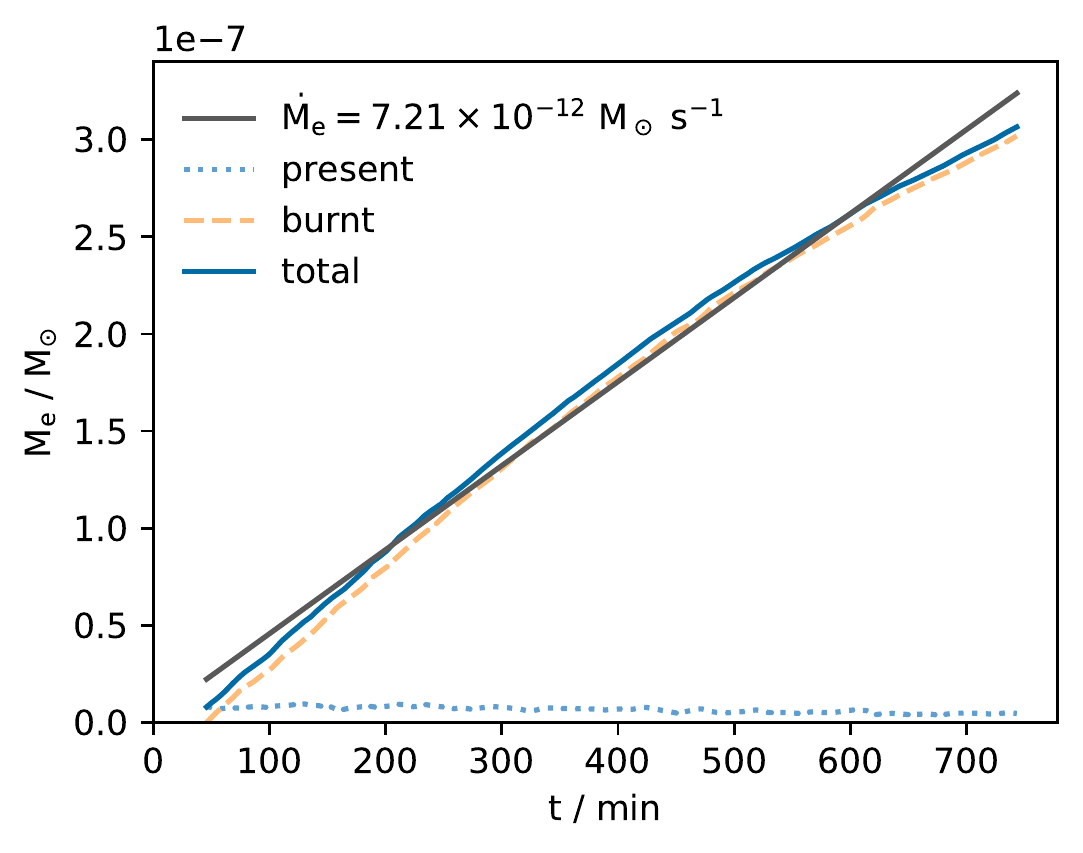}
    \includegraphics[width=\columnwidth]{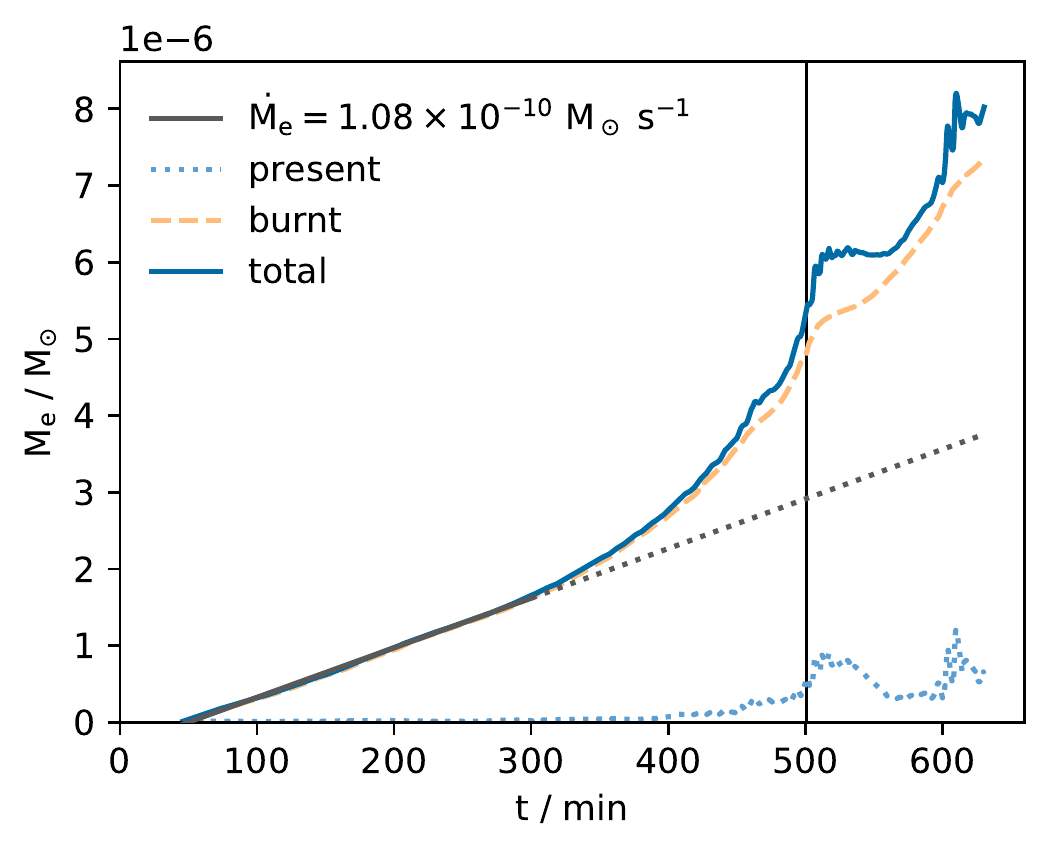}

  \centering
  \caption{The time evolution of the mass of the \cldfluid{} being
          entrained and burned within the convection zone since the start of the \mppnp{}
          post-processing models, $t = \unit{46}{min}$. The top panel is from \code{N15},
          the middle panel is from \code{N16} and the bottom panel is from \code{N17}. The
          total lines is the instantaneous entrainment rate of the \cldfluid{} which
          is used in all advective and diffusive post-processing models. A linear fit of
          the entrainment rates, $\dot{M}_{\mathrm{e}}$, for each run is calculated over a
          quasi-static time interval corresponding to the solid line, while the dotted line
          is an extension of that fit. The black vertical line is at $t = \unit{501}{min}$
          when the images of \Fig{renderings} were rendered.}
  \label{fig:entrain}

\end{figure}

At the upper CB, the entrained \cldfluid{} follows the convective
downflows deep into the convection zone until it reaches the burning region
where the \nuclei{12}{C}$(\pt,\gamma)$\nuclei{13}{N} reaction occurs. To
determine the entrainment rates of \code{N15}, \code{N16} and \code{N17}, a
method, which is described in more detail in \citet{Andrassy:19} and
\citet{Denissenkov:19}, was used. To summarize, the total mass of the
\cldfluid{} that was entrained is composed of the mass that has been
burned and the mass that is present within the convection zone with an upper
boundary radius of $r_\mathrm{ub}$. To determine the mass of the \cldfluid{}
within this convection zone its density is integrated to within
\unit{0.5}{Mm} of the formal CB of the \mppnp{} post-processing simulations
(see \Section{modelConstraints}). This offset is approximately the average scale
height of the rms tangential velocity gradient at the boundary, $H_{v_{\perp},
\mathrm{ub}} = (\partial \: \mathrm{ln} \: v_{\perp} / \partial r)^{-1}$, a
condition used in \citet{Jones:2017kc}, \citet{Andrassy:19}, and
\citet{Denissenkov:19}, for all RAWD simulations during their quasi-static
burning phases. The constant offset prevents major fluctuations in the
calculated entrained mass due to the very large concentrations of the
\cldfluid{} near the CB and prevents major changes in the velocity field
(global oscillation of shell H ingestion in \code{N15} and \code{N17}; see \Section{goshruns}) from providing
outlandish integration boundaries. This offset is extended to \unit{5}{Mm} for
\code{N17} during its global oscillation of shell H ingestion due to the major changes in the spherically averaged
\cldfluid{} concentrations near the majorly perturbed convective boundary. The
burnt material is calculated using the spherical profiles of density,
temperature and mass fraction to determine what the burning rate per unit volume
is. This burning rate is integrated in time to determine the amount burnt over a
\textit{dump}. The entrainment rates, mass burnt, and present material of the
\cldfluid{} within the CBs of \code{N15}, \code{N16} and \code{N17} are
shown in \Fig{entrain}. A linear fit of the entrainment rate of \code{N15},
\code{N16}, and \code{N17} over their quasi-static burning phases are
$\dot{M}_{\mathrm{e}} = 1.06 \times 10^{-11}\,M_{\odot}$\,s$^{-1}$,
$\dot{M}_{\mathrm{e}} = 7.21 \times 10^{-12}\,M_{\odot}$\,s$^{-1}$, and
$\dot{M}_{\mathrm{e}} = 1.08 \times 10^{-10}\,M_{\odot}$\,s$^{-1}$,
respectively. \code{N15} and \code{N17}'s entrainment rates become non-linear
and increases significantly around $t = \unit{300}{min}$ (\Fig{goshrprofs}).
Even within the linear regime the entrainment rate of \code{N15} is $\approx
10\%$ larger than \code{N16}'s though the scale of this difference is consistent
with the entrainment convergence results in Fig.~17 of \citet[][]{woodward15}.

Throughout the entirety of the \code{N16} simulation, the burning and
entrainment of H maintains a quasi-static state resulting in the linear
growth of the total entrained material. The distribution of $X_{\mathrm{H}}$ is
plotted on spherical shells near the upper CB, \unit{23}{Mm}, and well within
the H burning region, \unit{14.5}{Mm}, in \Fig{mollweide}. The corresponding
radial velocity field at those radii is shown in \Fig{mollweide}. The
distribution of the $X_{\mathrm{H}}$ at \unit{14.5}{Mm} in conjunction with the
radial velocity distribution shows that that the upflows are essentially H-free
while the downflows are H-rich. As the H-rich material moves through the H
burning region, the downflow material is rapidly burned until it is H-free.
Eventually this material will turn around and move in the upflows entirely
H-free. The H-free material is advected and mixed with the H-rich material as it
is advected towards the upper boundary where it still maintains very small mass
fractions ($X_{\mathrm{H}} \approx 10^{-6}$) at \unit{23}{Mm}, $\approx
\unit{0.8}{Mm}$ from the upper CB (\Tab{ppmstarmodels}). The H-rich material
($X_{\mathrm{H}} \approx 10^{-3}$) is advected into the convection zone in the
downflows near the CB.

\subsubsection{A global oscillation of shell H ingestion in \code{N15} and \code{N17}}
\label{sec:goshruns}

\begin{figure}

  \includegraphics[width=\columnwidth]{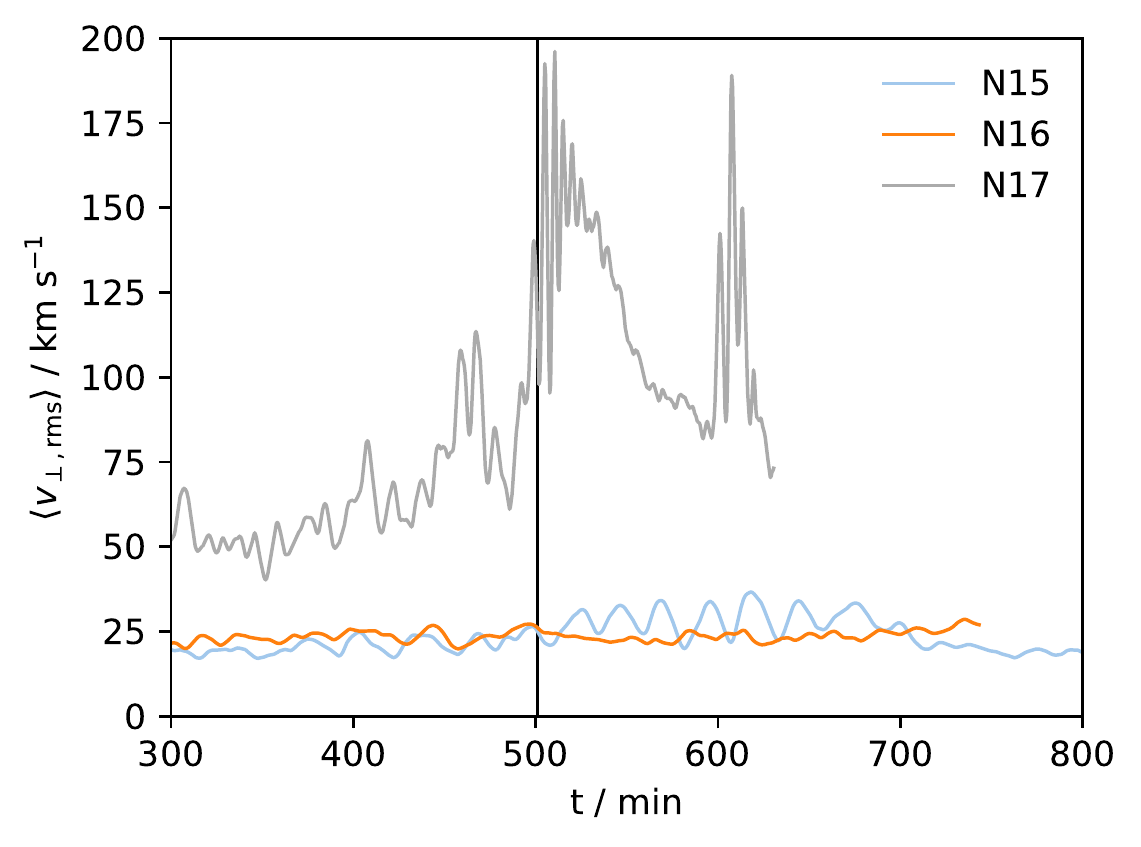}
  \centering
  \caption{The average of the rms tangential velocity between 1 and \unit{2}{Mm}
          below each simulations top Schwarzschild boundary as described in
          \Section{convbound}. The very large oscillations in the tangential velocity
          in \code{N17} indicate a global oscillation of shell H ingestion is occurring which causes the entrainment rate to
          increase by up to a factor of 80 (see \Fig{entrain}) during these oscillations.
          Much milder oscillations can be seen in \code{N15} between 500 and
          \unit{700}{min} with a modest factor of 3 increase in its entrainment rate. The
          black line refers to the time when the images of FV in \Fig{renderings} were
          rendered.}
  \label{fig:goshrprofs}

\end{figure}

\begin{figure*}

  \centering
  \includegraphics[width=\textwidth]{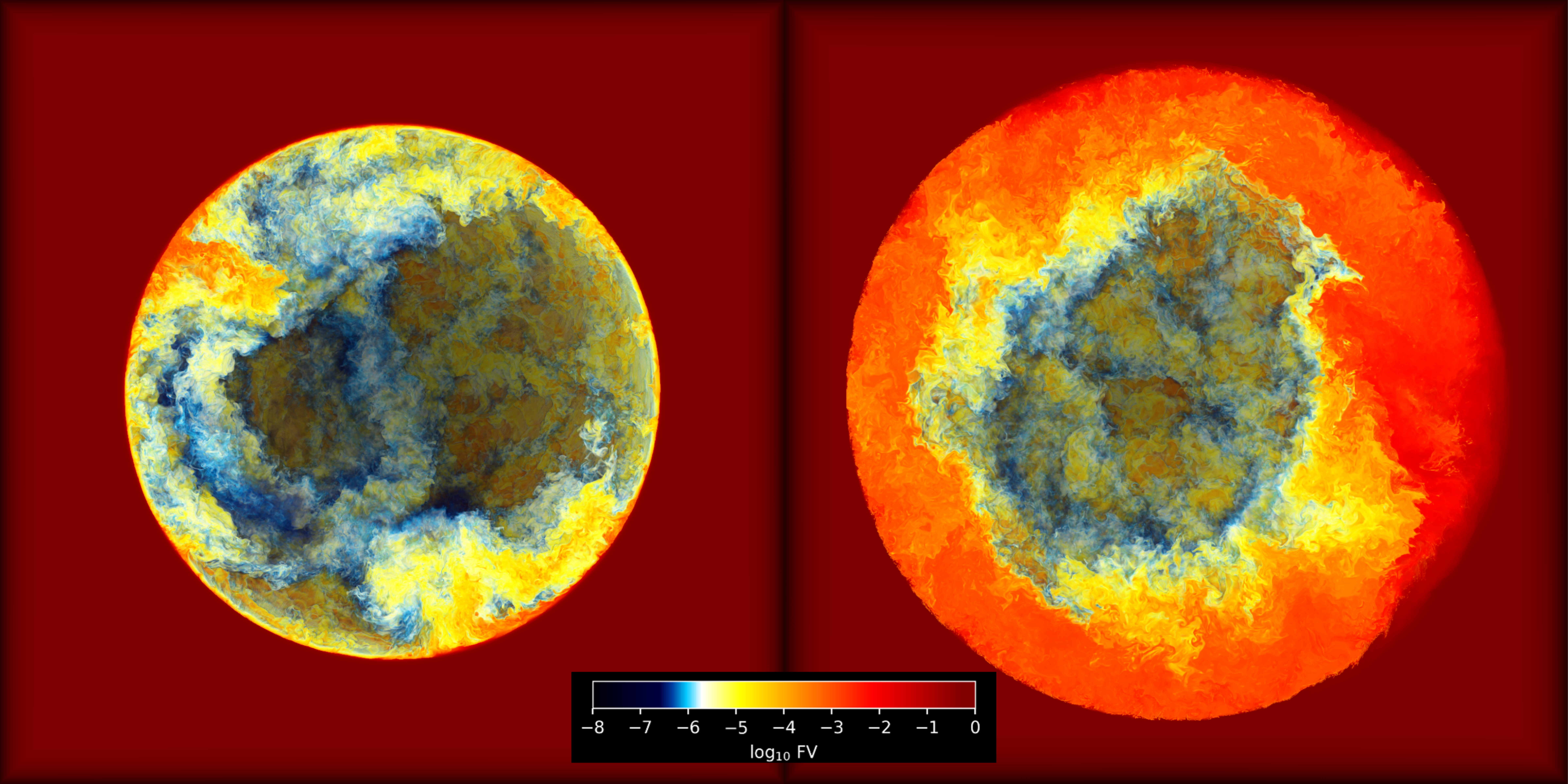}
  \caption{Both panels show a rendering of the fractional volume FV of
          the material in the stable layer at $t = \unit{501}{min}$. The C-rich \airfluid{} is
          transparent. The left panel is from \code{N16}, while the right panel is from
          \code{N17}. The entrainment rate of H rapidly increases when a global oscillation of shell H ingestion instability
          occurs in \code{N17}. The hydrodynamic feedback from the energy released from
          burning causes even more H to be entrained and leads to an unstable runaway in
          \code{N17} but not in \code{N15}. This is contrasted with the quasi-static
          entrainment and burning in \code{N16} (\Fig{entrain}).}
  \label{fig:renderings}

\end{figure*}

Directly from the entrainment rates of each run in \Fig{entrain}, the
entrainment rates begin to increase substantially around $t = \unit{300}{min}$
in both \code{N15} and \code{N17} and continues until $t \approx
\unit{700}{min}$. \code{N17}'s entrainment rate increases by a factor of up to
80 during this time compared with its quasi-static rate, while \code{N15}'s
entrainment rate increases by a factor of up to 3. The cause of these bursts of
entrainment are due to the collision of opposing horizontal flows forcing
significant amounts of H-rich material to be entrained in downdrafts where it
will eventually burn and feedback energy into the flow~\citep{Herwig:2014cx}.
The magnitude of the horizontal oscillations of these flows is clearly seen in
\Fig{goshrprofs} where the rms of the tangential velocity rapidly increases and
decreases within convective turn over timescales. With the same heating rate as
\code{N15}, \code{N16} does not undergo a global oscillation of shell H ingestion at any point during its entire
simulation which includes the full duration of the global oscillation of shell H ingestion experienced by
\code{N15}.

The consequences of the global oscillation of shell H ingestion between \code{N15} and \code{N17} differ
drastically due to the differences in the amount of H ingested. The weak global oscillation of shell H ingestion of
\code{N15} does not entrain enough H to sustain it for a long period of time and
thus it dissipates after \unit{200}{minutes} of large scale oscillations. The
burning of this additional entrained H is done such that there is very little
build up of H within the convection zone (\Fig{entrain}) during the global oscillation of shell H ingestion. The
global oscillation of shell H ingestion does not increase the average tangential velocity significantly over its
duration and a global oscillation of shell H ingestion does not occur again throughout the \unit{1600}{minute} long
simulation. Conversely, \code{N17}'s global oscillation of shell H ingestion causes significantly higher
entrainment of H which is built up within the convection zone. This can be seen
in the rendering of the fractional volume, FV, at $t = \unit{501}{min}$ in
\Fig{renderings}. The global oscillation of shell H ingestion subsides briefly after the large build up of H and
burns most of the H within the convection zone and then begins again at $t =
\unit{600}{min}$. The simulation is ended soon after the global oscillation of shell H ingestion continues again as
the expansion of the convection zone has approached where the outer boundary
condition is applied.

\subsection{\mppnp{} post-processing model constraints}
\label{sec:modelConstraints}

Due to the initial transient at the start of any \ppmstar{} simulation, the
\mppnp{} post-processing models do not begin until well after this transient has
finished. This was chosen to be at $\unit{46}{minutes}$ for
\code{N16} and \code{N17}. The advective mixing models are initialized with an
$X_{\mathrm{H}}$ profile in the up- and downstreams that is equivalent to the
spherically averaged $X_{\mathrm{H}}$ profile from the \ppmstar{} simulation.
The diffusive mixing model is initialized with the spherically averaged
$X_{\mathrm{H}}$ profile from the \ppmstar{} simulation. For the advective
post-processing it is not expected that the up- and downstreams would have
equivalent $X_{\mathrm{H}}$ profiles as seen in \Fig{mollweide}. While running,
the initial guess of the equivalent profiles in the up- and downstreams is
quickly changed to the model's preferred profile which is asymmetric (see
\Fig{advectionprofiles}). This takes around 3 convective turn over timescales
and causes comparisons with \ppmstar{} profiles to be inaccurate during this
initialization period. For this reason, all \mppnp{} post-processing models
repeat the very first time step, with the entrainment rate at that point in
time, for $\approx 3$ convective turn over timescales so that they establish
their own quasi-static profile.

The \ppmstar{} simulations output the \textit{briquette}, and any other data
type, on a \textit{dump} basis. The advective mixing models have their time
steps being limited by a Courant condition, \eq{courant}, which requires on the
order of 20-30 time steps being taken for every single \textit{dump} of
\code{N16}\footnote{\ppmstar{}'s Courant condition is based on the speed of
sound}. Although the implicit solving of the diffusion equation used by \mppnp{}
is not limited to a time step criterion like the advective mixing model, all
post-processing models for a given run, regardless of mixing or network, use the
advective mixing models Courant condition limited time steps. Both the
advective and diffusive post-processing models, for a given 3D
\ppmstar{} run, use the same spherically averaged radial $\rho$, $T$ and $v_{\mathrm{rms}}$
profiles at a \textit{dump}. Although the stratification and velocity profiles do change
throughout a \textit{dump} it is negligible as the convective turn over time is
$\approx \unit{1100}{s}$ for \code{N16}, while the dump interval is $\approx
\unit{27}{s}$.

As discussed in \Section{convbound}, the CBs in 3D hydrodynamic simulations are
not a sharp and spherically symmetric boundary as typically interpreted with the
Schwarzschild criterion. The minimum of the spherical average of $\partial
v_{\perp} / \partial r$, which can define the location of the CB, could be used
to determine a time dependent CB for the post-processing models. However, with
this CB varying at each \textit{dump} in the Eulerian coordinates it is also
varying in its mass coordinates. Because the advective mixing model requires
that the mass of the individual cells to remain constant across all time
(\Section{advectmodel}) this boundary determination can cause the upper and
lower boundaries to sporadically move between cells on a \textit{dump} basis.
During a global oscillation of shell H ingestion this boundary determination can lead to a top boundary well within
the convection zone, as it was determined before it began to global oscillation of shell H ingestion, due to the
large oscillations in the tangential velocities (see \Fig{goshrprofs}). To
simplify the definition of the CB and make it applicable to all runs, instead
the mass coordinates of the initial Schwarzschild boundaries are used to define
the convective boundaries for all post-processing models. The difference between
the two boundary criteria is only $\unit{0.43}{Mm}$ at $t = \unit{299}{min}$ for
\code{N16} (\Tab{ppmstarmodels}). Given that the determination of the tangential
velocity gradients are from the \textit{briquette} data, which is downsampled
from the \ppmstar{} simulation and only have a cell length of $\unit{0.18}{Mm}$,
and the fact that $\sigma_{r_{\mathrm{b}}, v_{\perp}} = \unit{0.46}{Mm}$ for
that averaged boundary the two boundary determinations are mostly in agreement.

The mass coordinates of the cell interfaces, which are constant for the whole
duration of the post-processing, are calculated by initially splitting the
convection zone into equally spaced radial shells in the Eulerian coordinates.
This makes the mass of individual cells to vary radially but this ensures that
the sampling of data from the hydro simulations is done at the cell resolution.
There are approximately $250$ cells for the post-processing models of \code{N16}
and \code{N17}. The number of cells in \code{N16} was reduced due to the
computational effort required when using a full network. As the \ppmstar{}
simulation evolves in time, the density, velocity and radius are interpolated to
the mass coordinates of the cell interfaces.

\begin{figure}

  \includegraphics[width=\columnwidth]{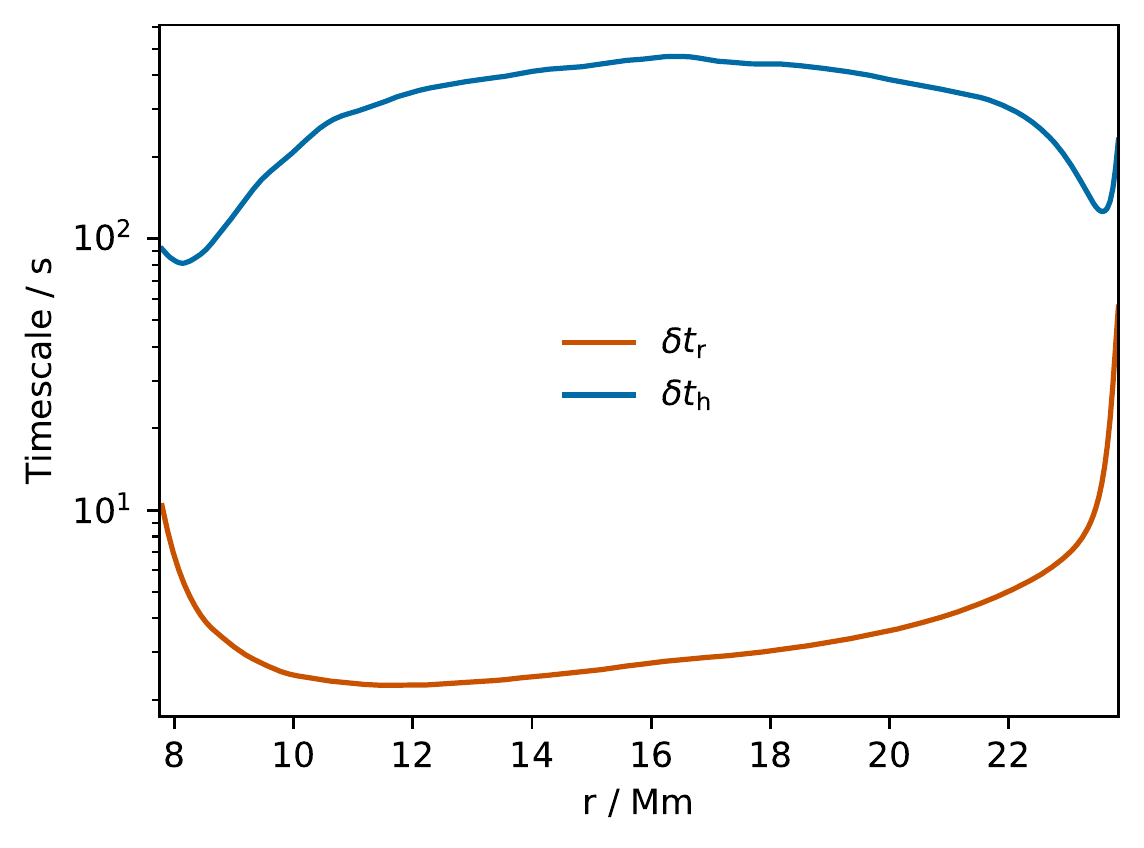}
  \centering
  \caption{The radial and horizontal mixing timescales as a function of radius
          within the convection zone at $t = \unit{299}{min}$ for run \code{N16}.
          Equations~\ref{eq:radialT} and~\ref{eq:horizontalT} define the radial and
          horizontal mixing timescales, respectively.}
  \label{fig:timescales}

\end{figure}

The stream model for \code{N16} was run from $t = \unit{46}{min}$ until the end
of that simulation at $t = \unit{744}{min}$ resulting in roughly 39
convective turn over times. The entrainment of the \cldfluid{} into the
convection zone of the post-processing models is taken directly from the time
dependent entrainment in \Fig{entrain}. The entrainment rates are small enough
that the fact that the mass of the convection zone does not change over the
length of the post-processing model is an accurate approximation to the
\ppmstar{} simulations. Integrating \code{N16}'s entrainment rate over the
length of its simulation results in a total of $3.1 \times 10^{-7}\,M_{\odot}$
of \cldfluid{} being entrained. The cell with the smallest mass within the
advective post-processing models of \code{N16} is $2.6 \times 10^{-6}\,M_{\odot}$.

\subsection{$\nuclei{12}{C}(\pt,\gamma)\nuclei{13}{N}$-only post-processing models}
\label{sec:c12pg}

\subsubsection{Diffusive post-processing models}
\label{sec:Dc12pg}

\begin{figure}
 \includegraphics[width=\columnwidth]{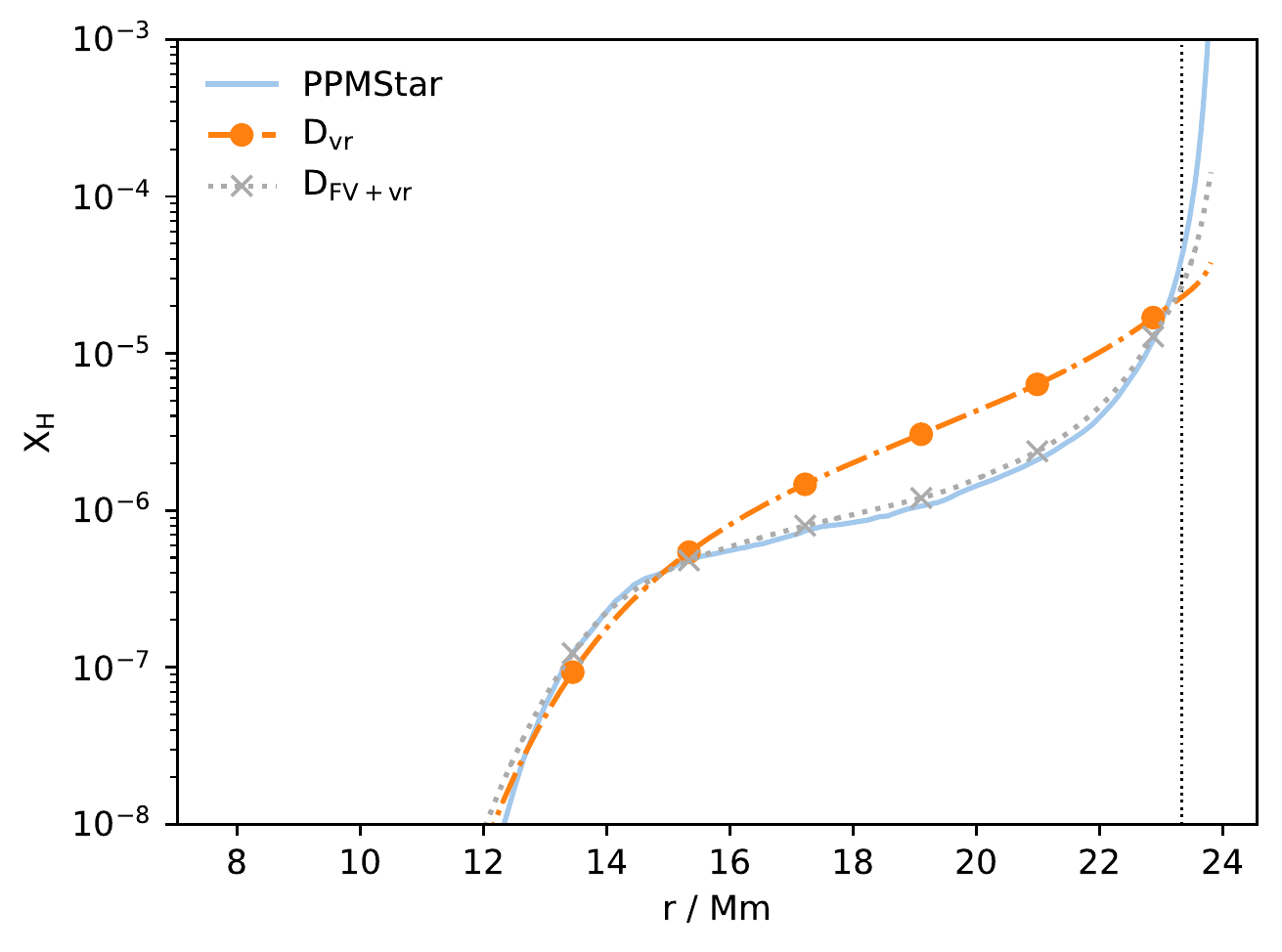}
  \caption{The $X_{\mathrm{H}}$ profiles from \code{N16} and from
          post-processing models \code{mp9}, using $D_\mathrm{vr}$ according to the recipe
          provided in \citet{Jones:2017kc}, and \code{mp1}, using $D_\mathrm{FV+vr}$ which
          inverts the diffusion equation using the spherically-averaged $X_\mathrm{H}$
          profiles of \ppmstar{} as described in \Section{Dmethod}.}
  \label{fig:ppmstar-d-compare}
\end{figure}

Using spherical averages of the $X_{\mathrm{H}}$ and rms of the radial velocity
from \ppmstar{} the diffusion coefficients $D_{\mathrm{vr}}$ and
$D_{\mathrm{FV+vr}}$ are computed on a \textit{dump} basis. Using these
diffusion coefficients in \mppnp{} with only the $\nuclei{12}{C}(\pt,\gamma)\nuclei{13}{N}$
reaction, the post-processing models of \code{mp1} and \code{mp9} are compared
with \ppmstar{} in \Fig{ppmstar-d-compare}. Unsurprisingly, the model using the
$D_{\mathrm{FV+vr}}$ matches the \ppmstar{} profile very closely except right at
the top CB. This is due to the accuracy of estimating the strength of the mixing
and the calibration of the entrainment rates. By using the $D_{\mathrm{vr}}$
diffusion coefficient the mixing near the convective boundary is largely
overestimated and the mixing near the middle is underestimated. Even though the $|r
- r_{0}|$ term in \eq{dvr} decreases as the boundary is being approached,
$D_{\mathrm{vr}}$ does not fall off as quickly as the $D_{\mathrm{FV+vr}}$
profiles suggest in \Fig{DvrDVF-compare}.

\subsubsection{Advective post-processing models}
\label{sec:Advc12pg}

\begin{figure}

  \includegraphics[width=\columnwidth]{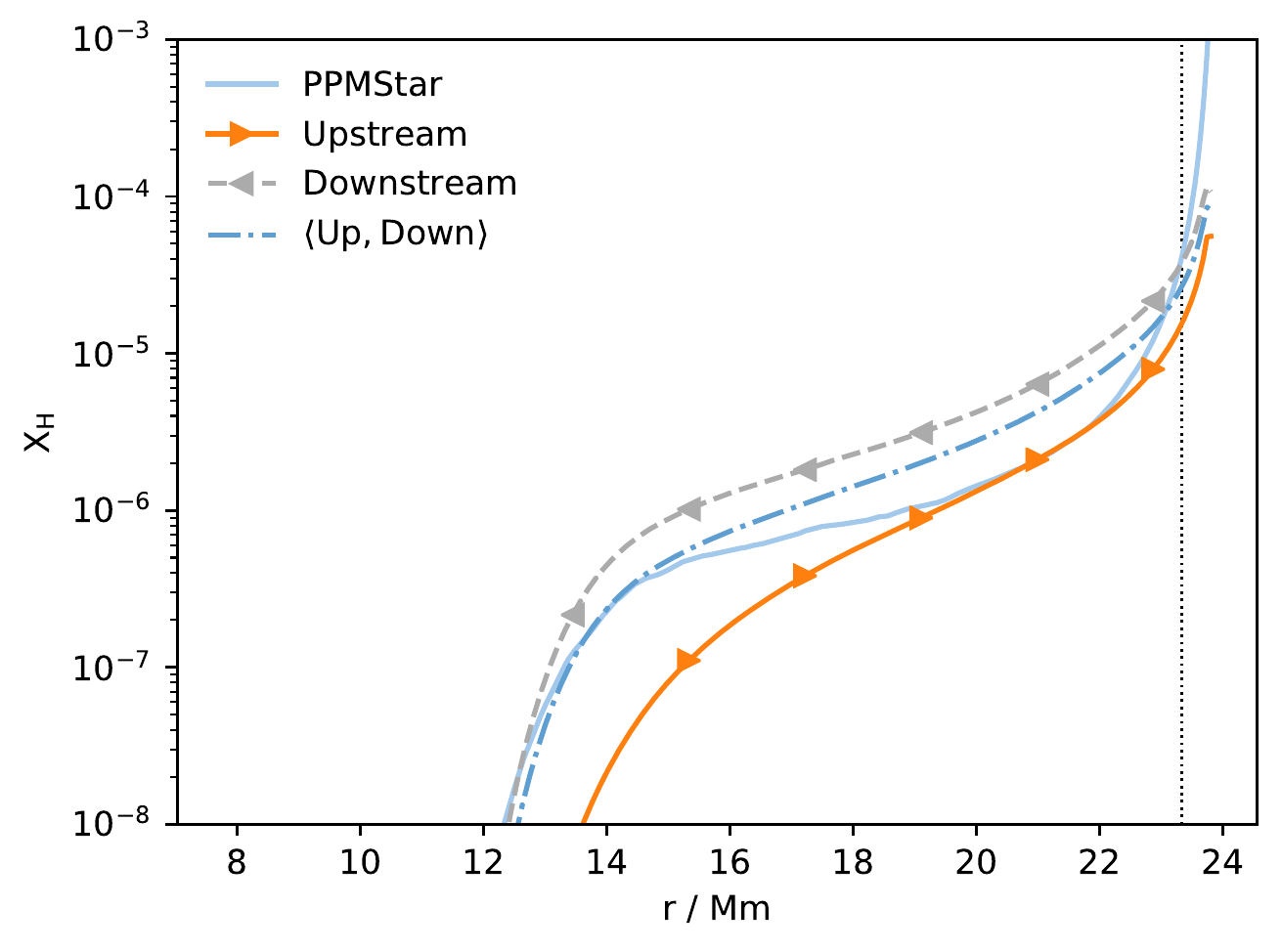}
  \includegraphics[width=\columnwidth]{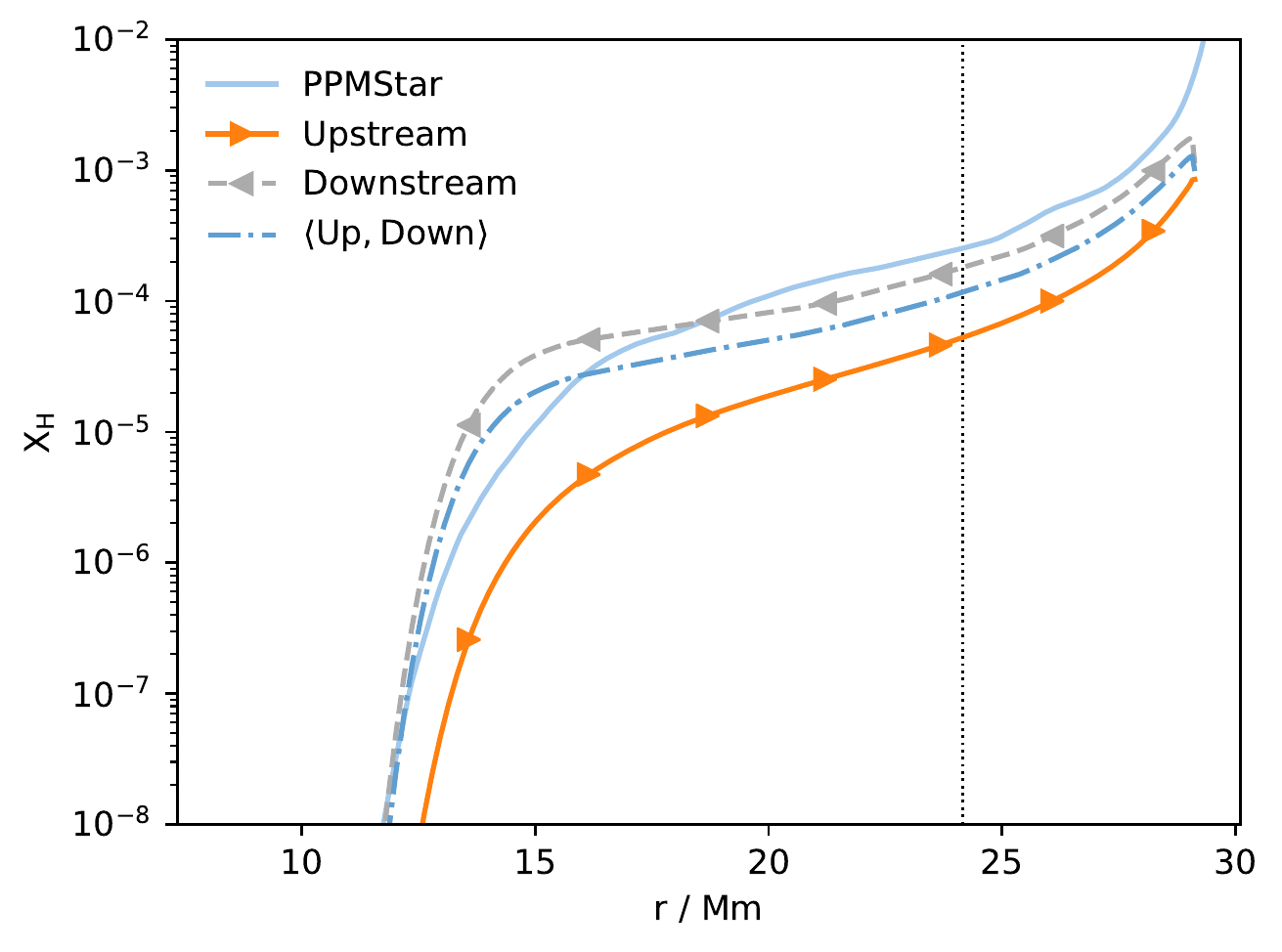}

  \centering
  \caption{The top panel is the $X_{\mathrm{H}}$ profiles from \code{N16} and
          the up- and downstreams of \code{mp2} at $t = \unit{299}{min}$. The spherical
          average of the up- and downstreams is $\langle \mathrm{Up}, \mathrm{Down}
          \rangle$. The bottom panel plots the same quantities for \code{N17} and the
          post-processing model \code{mp4} at $t = \unit{501}{min}$, during the global oscillation of shell H ingestion. The
          black vertical dotted line is where the integration of the \cldfluid{} is
          stopped for calculating the entrainment rates in \Fig{entrain}.
        }
  \label{fig:advectionprofiles}

\end{figure}

From \FigTwo{power-spectrum}{lspectrum} the power spectra of \code{N16} shows
dominant large-scale modes at the middle of the convection zone as well as a
more flat spectrum across many scales when near either CB. With the smaller
modes contributing significantly to the power near the CBs the mixing between
the two streams is expected to be more efficient there than the middle of the
convection zone. Applying \eq{horizontalT}, the horizontal and radial timescales
for run \code{N16} are shown in \Fig{timescales}. The horizontal timescale is
typically over an order of magnitude larger than the radial timescale suggesting
inefficient mixing, even at the CBs. There is still significant power at the
largest modes near the CBs leading to small changes in the horizontal
timescales across the entire convection zone. The lack of efficient horizontal
mixing is especially apparent in the middle of the convection zone where the up-
and downstreams are very isolated from each other as seen in
\Fig{advectionprofiles}. The upstream is carrying nearly H-free material towards
the top of the convection zone, while the downstream is carrying H-rich material
directly to the burning region. This distribution of H-free fluid in the
upstream and H-rich fluid in the downstream is also validated by the 3D hydro
simulations as seen in \Fig{mollweide}.

The spherical average of the X$_{\mathrm{H}}$ profiles from the two streams
closely follows the spherical average of \code{N16} in most regions. However, at
the very top of the convection zone it is underestimated, while between 16 and
$\unit{22}{Mm}$ it is overestimated, similarly to the results of the diffusive
mixing with $D_{\mathrm{vr}}$ (\Fig{ppmstar-d-compare}). One component of this
is likely due to an underestimation of the horizontal mixing at the top of the
convection zone. The \textit{briquette} data used to compute the spectra of the
radial velocity has a resolution that is a factor of 4 smaller than the run's
grid resolution. This downsampling involves averaging which significantly
dampens the power in any short wavelength modes, directly increasing the
horizontal mixing timescale and leading to less efficient mixing. This results
in much more efficient radial transport of species allowing for the sliver of
very H-rich material at the top of the convection zone to immediately advect
downwards rather than be constantly mixed between the two streams.

\begin{figure}

  \includegraphics[width=\columnwidth]{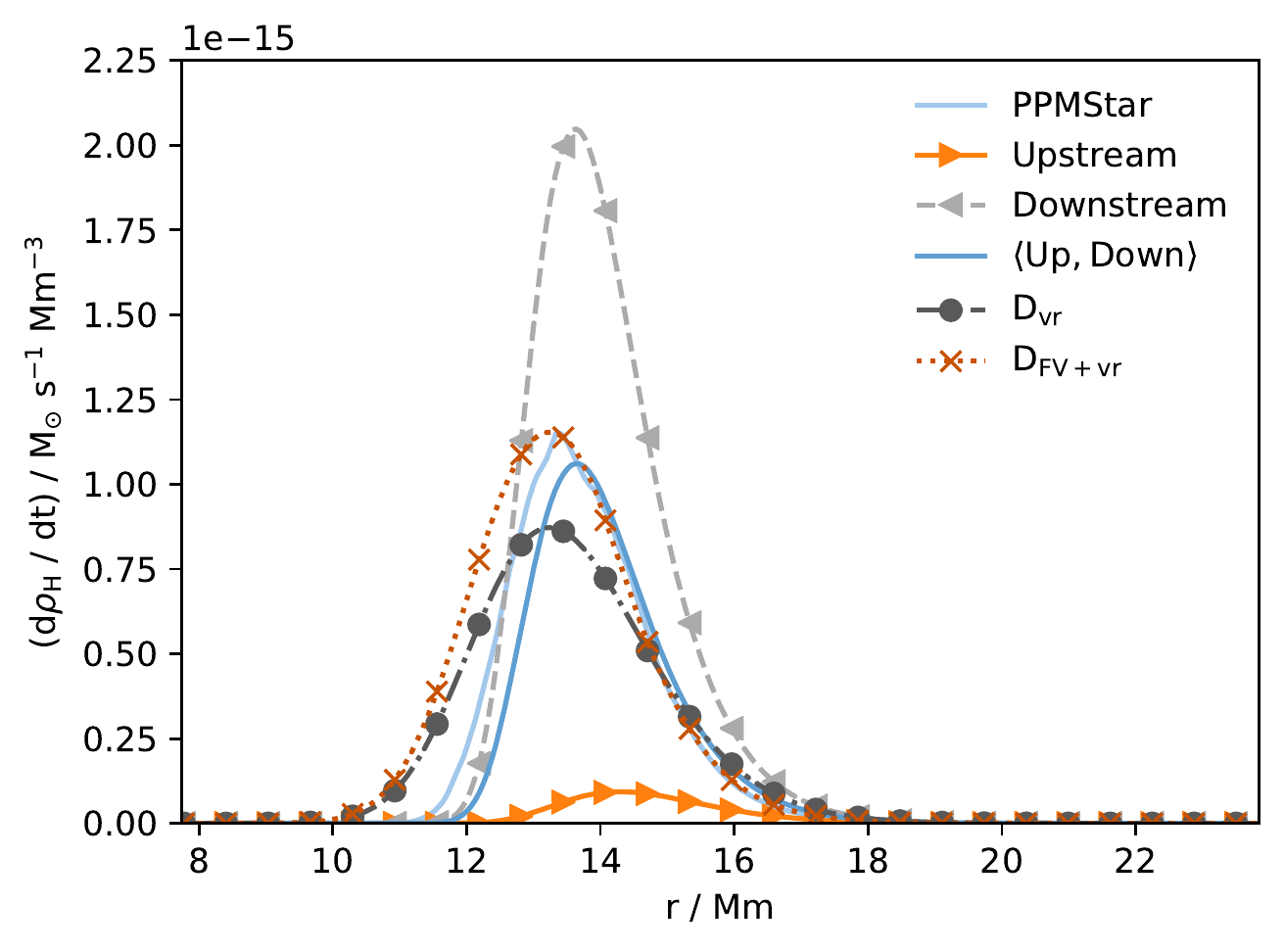}
  \centering
  \caption{The burning rate of H per volume as a function of radius
    for \code{N16}, an advective post-processing model, \code{mp2}, and the 
    diffusive post-processing models \code{mp1} and \code{mp9}. The $\langle
    \mathrm{Up}, \mathrm{Down} \rangle$ profile is calculated with the
    spherical average of the streams to estimate the burning rate,
    similarly to the estimates from spherically averaged data of
    \ppmstar{}. The \nuclei{12}{C}$(\pt,\gamma)$\nuclei{13}{N}
    reaction is very sensitive to temperature and it does not burn any
    H until about $\unit{18}{Mm}$ with $\mathrm{T9} = 0.11$. All of
    the H is burned well before it reaches the bottom of the
    convection zone.}
  \label{fig:burningrate}

\end{figure}

To better understand the physics of the convective-reactive nucleosynthesis in the RAWD model, 
it is useful to find out where the H-burning is
predominately occurring within its He shell. The H burning
rate at a single time step from the \ppmstar{} simulations and the advective
post-processing model, \code{mp2}, are estimated in \code{N16} with the
techniques discussed in \Section{convbound} and is shown in \Fig{burningrate}.
The advective post-processing model is burning the H a few cells above where
\code{N16} is burning H. This could suggest that the distribution of flow
velocities that are present in \ppmstar{} simulations modifies where the bulk of
the burning takes place within a convection zone. With the fact that there is no
significant accumulation of H throughout the advective post-processing model it
is roughly in a quasi-static state in which it is burning all of the H that it ingests.

\subsection{Full network post-processing and comparison with observations}
\label{sec:fulllnetworkresults}

\begin{figure*}

  \begin{minipage}{.49\textwidth}
    \centering
    \includegraphics[width=\columnwidth]{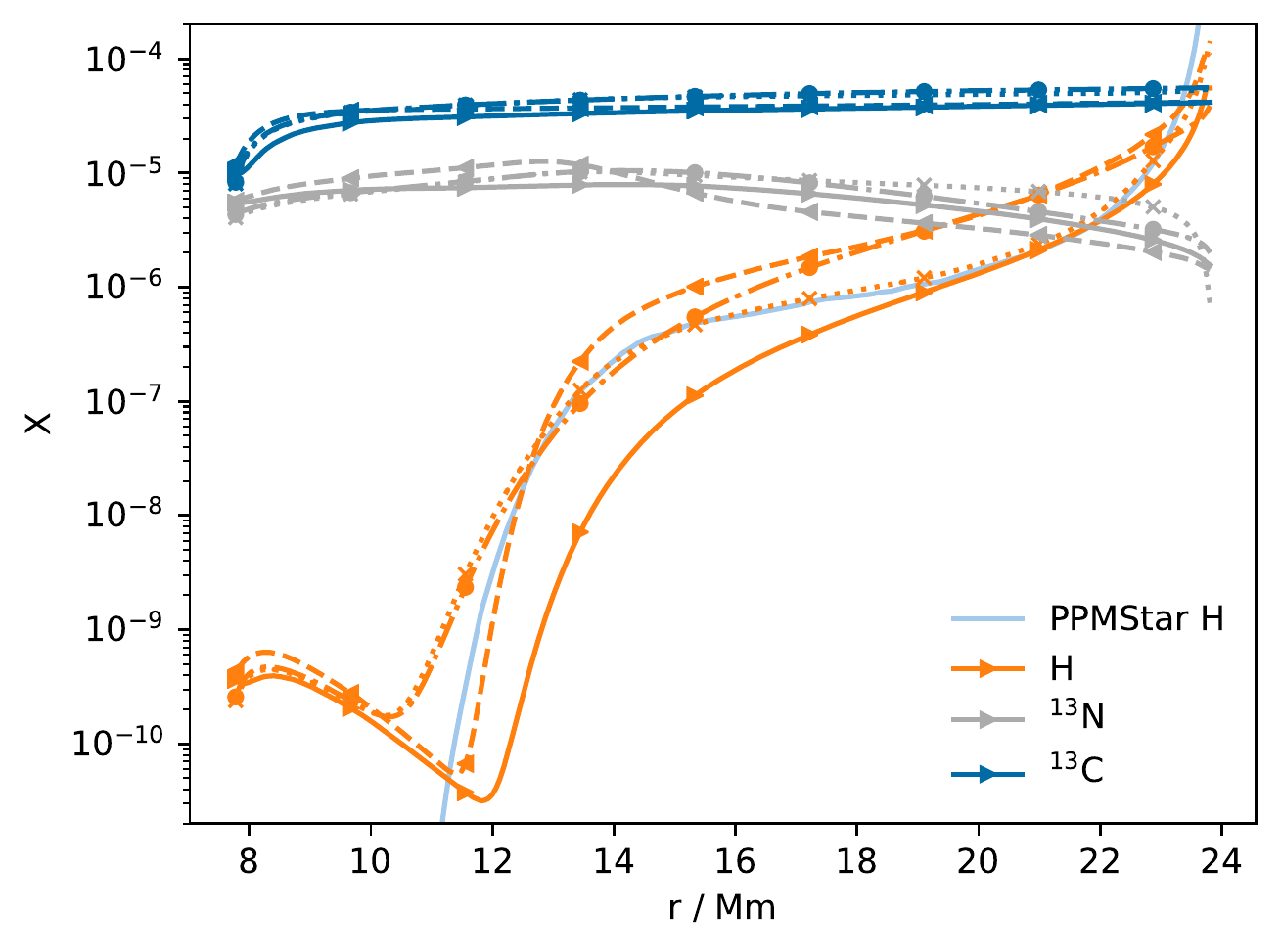}
  \end{minipage}
  \begin{minipage}{.49\textwidth}
    \centering
    \includegraphics[width=\columnwidth]{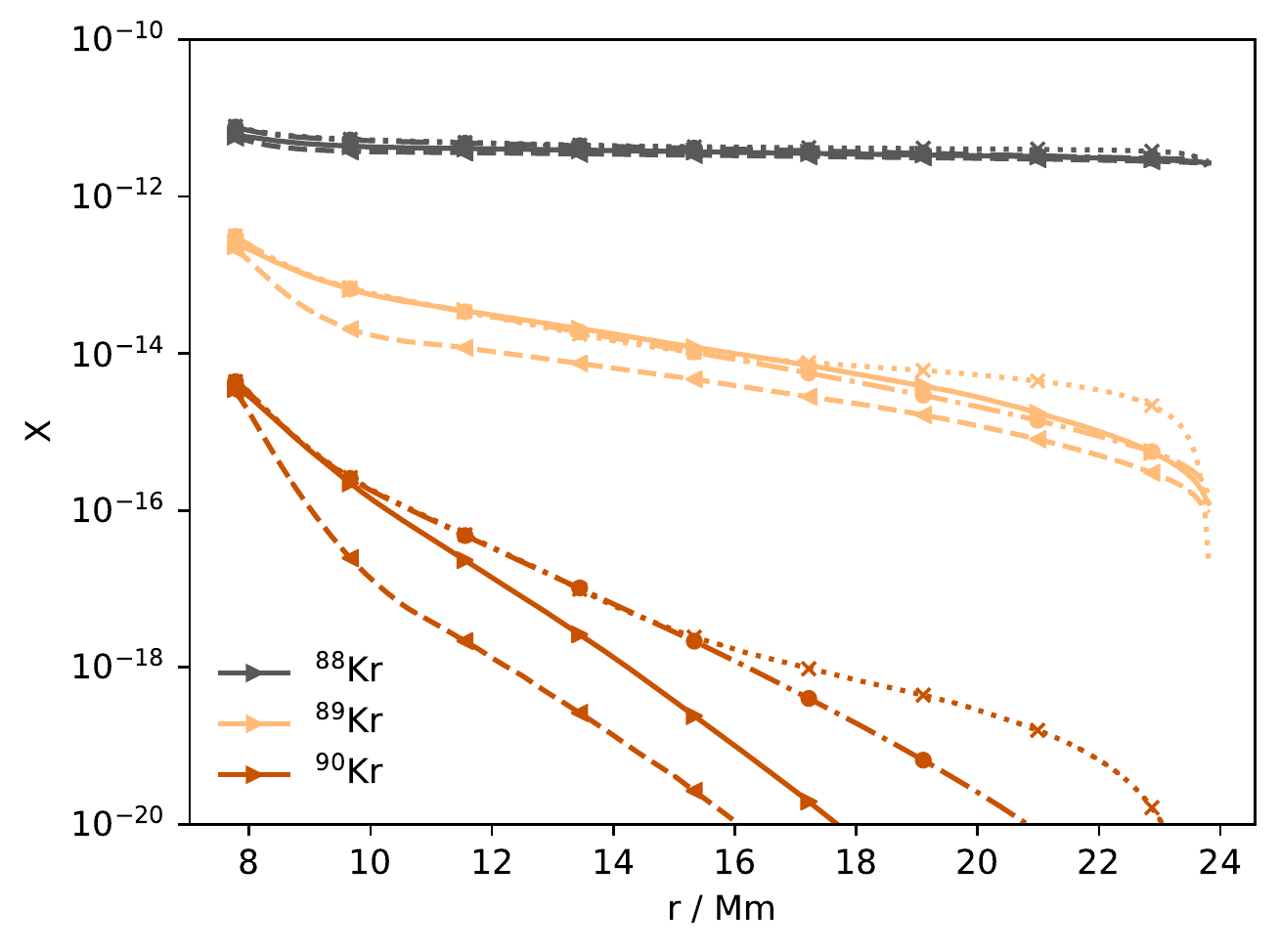}
  \end{minipage} \\

  \begin{minipage}{.49\textwidth}
    \centering
    \includegraphics[width=\columnwidth]{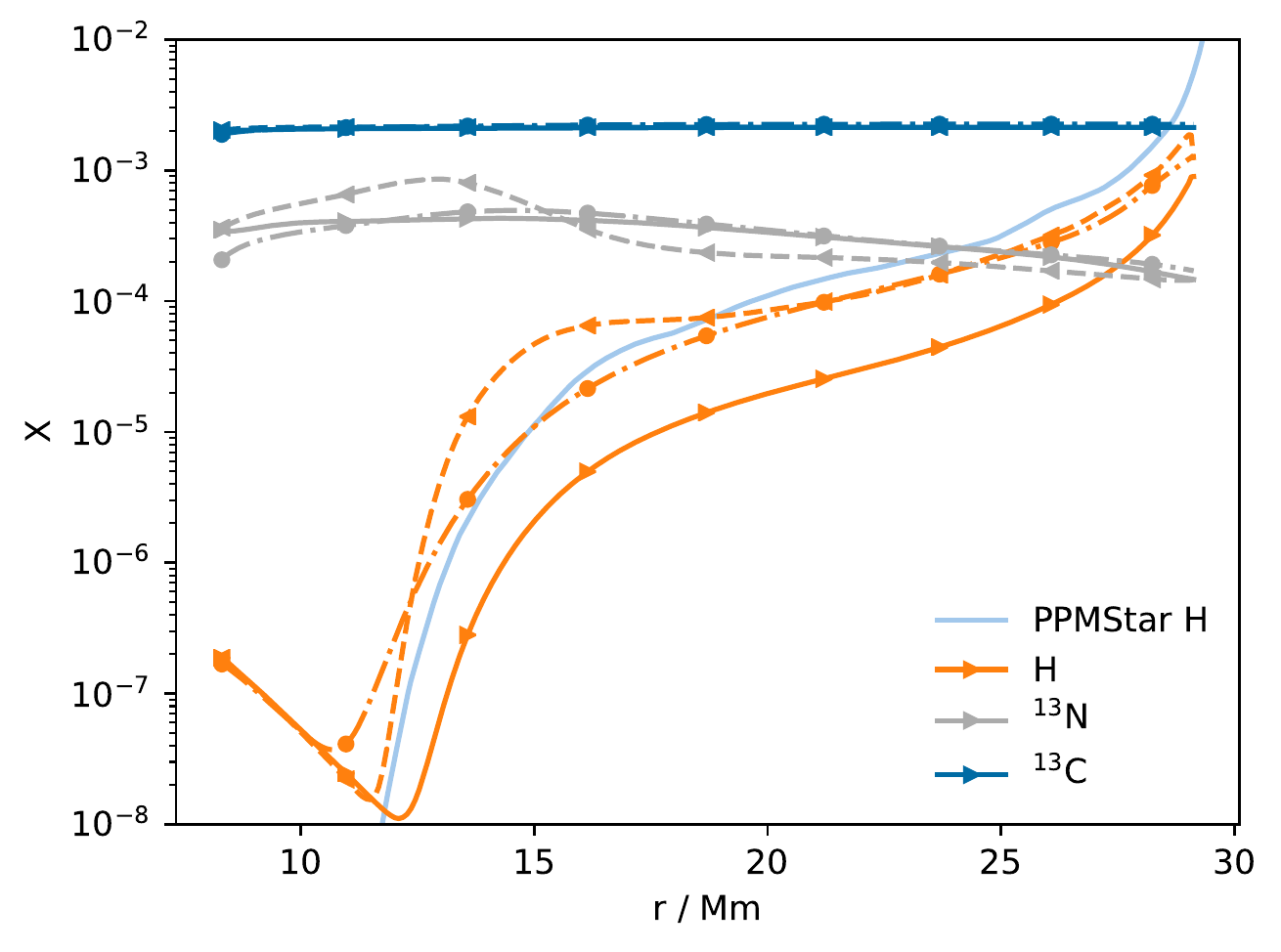}
  \end{minipage}
  \begin{minipage}{.49\textwidth}
    \centering
    \includegraphics[width=\columnwidth]{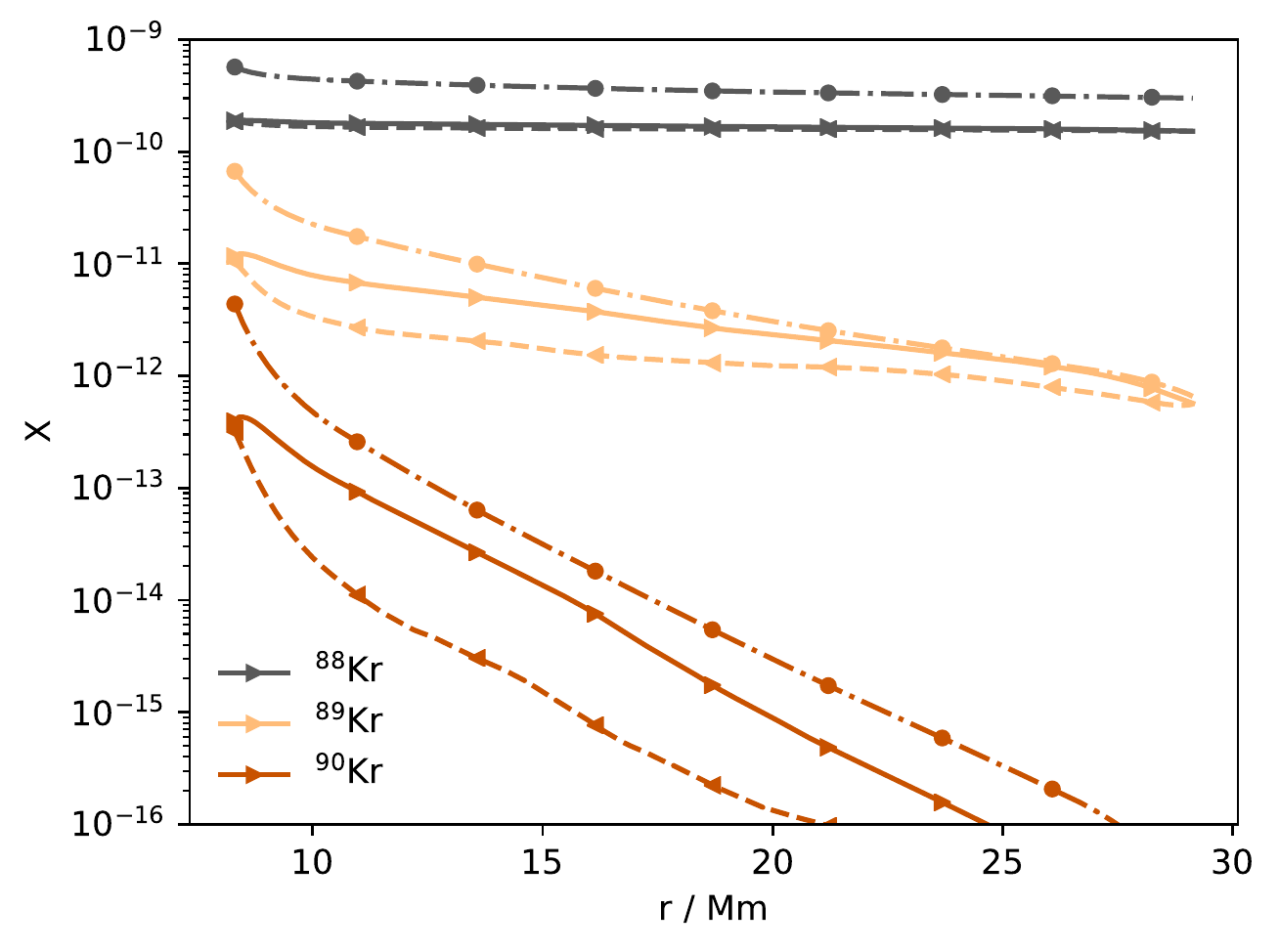}
  \end{minipage}
  
  \caption{The top row contains panels using data from the post-processing
          models of \code{mp5}, \code{mp6} and \code{mp10} (\code{N16}) at $t =
          \unit{299}{min}$, while the bottom row contains panels using data from the
          post-processing models of \code{mp7} and \code{mp8} (\code{N17}) at $t =
          \unit{501}{min}$. In each row, the left panel contains line profiles of the mass
          fractions of H, \nuclei{13}{N}, and \nuclei{13}{C}, while the right panel
          contains line profiles of the mass fractions of various Kr isotopes. These
          isotopes are \nuclei{88}{Kr}, \nuclei{89}{Kr}, and \nuclei{90}{Kr} which have
          half lives of $\unit{2.84}{hr}$, $\unit{3.18}{minute}$ and $\unit{32.3}{s}$,
          respectively. The mixing models are distinguished based on line style and glyphs
          as was done in \FigTwo{ppmstar-d-compare}{advectionprofiles}. $D_{\mathrm{vr}}$
          profiles are dash-dotted with circles, $D_{\mathrm{FV+vr}}$ profiles are dotted
          with cross', the upstream profiles are solid with a triangle pointing towards
          higher radii and the downstream profiles are dashed with a triangle pointing
          towards lower radii.}

  \label{fig:fnet}
\end{figure*}

\begin{figure}

  \includegraphics[width=\columnwidth]{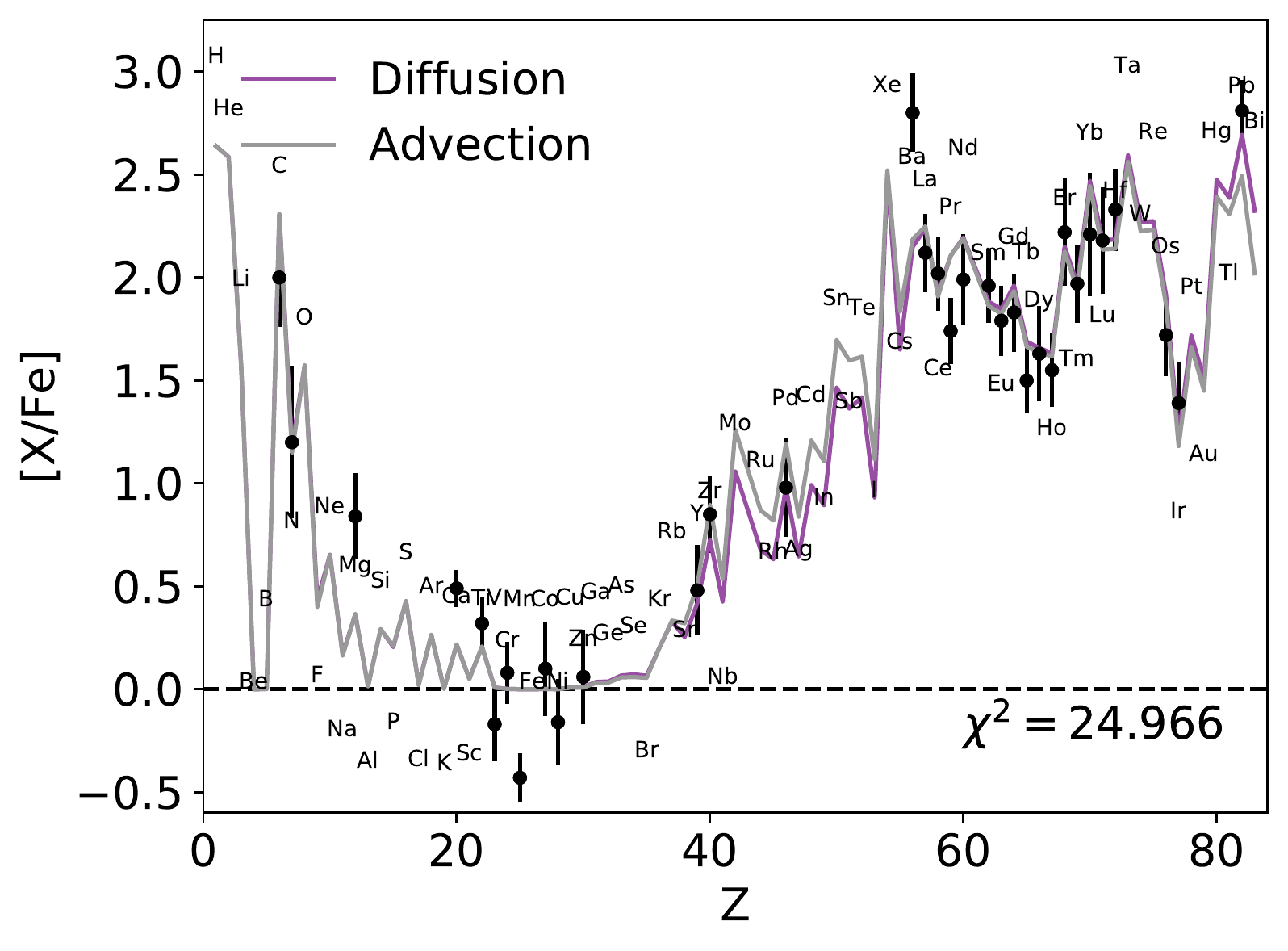}\\
  \includegraphics[width=\columnwidth]{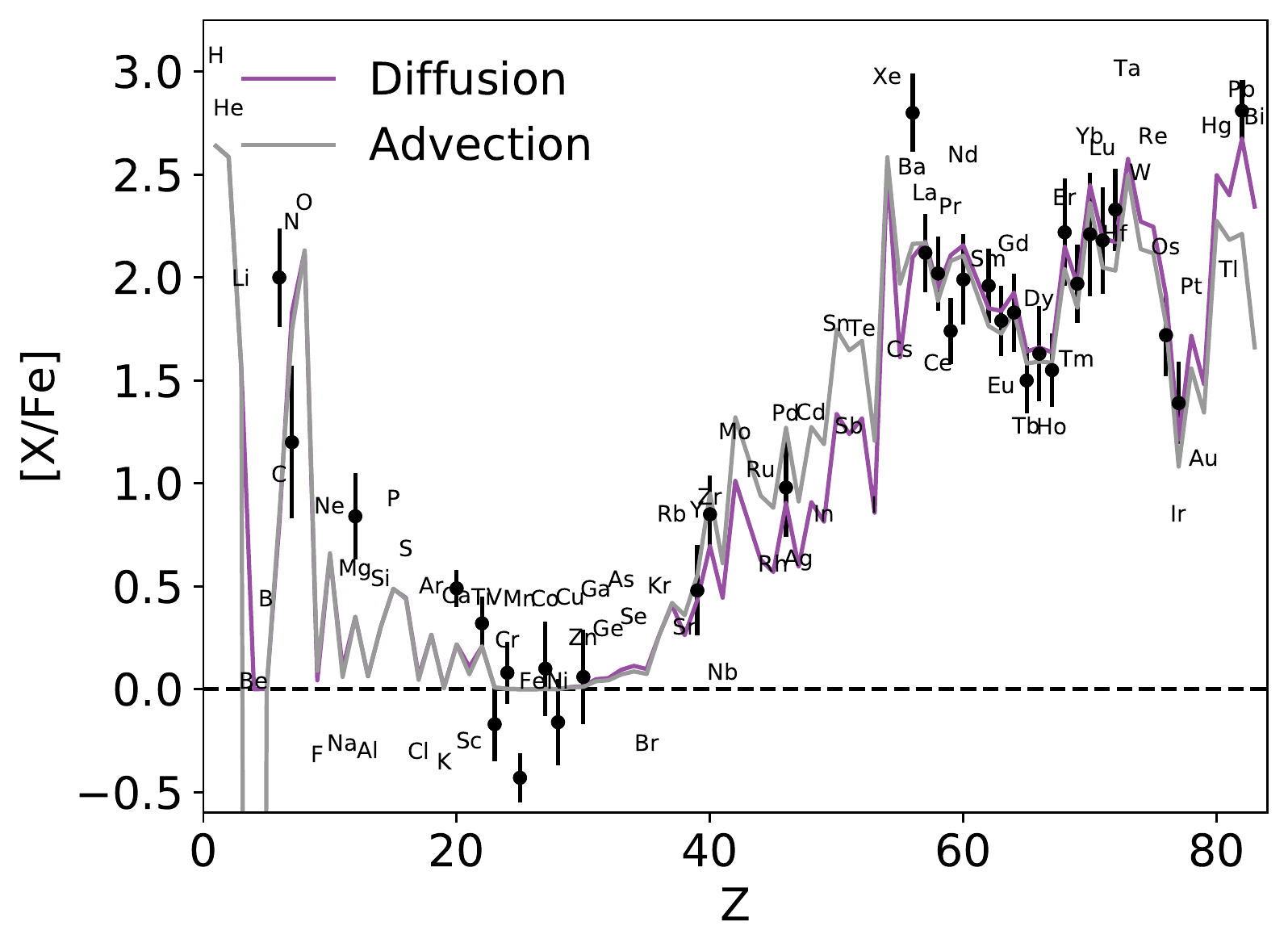}
  \centering
  \caption{Both panels show an elemental abundance distribution of the best-fit
          time for a diffusive post-processing model (\code{mp5} at 0.059 yr for \code{N16} and
          \code{mp7} at 0.017 yr for \code{N17}) with star CS31062-050~\citep{aoki:02c, johnson:04} as
          well as the appropriate advective post-processing model at the same simulation
          time (\code{mp6} for \code{N16} and \code{mp8} for \code{N16}). The top panel
          plots the models from run \code{N16} with the lower $N_{\mathrm{n}}$ thus taking
          longer to reach the \nuclei{}{Pb} elemental abundance than the bottom panel
          which uses models of \code{N17}. The abundances are scaled by the solar
          abundances from \citet{Asplund2009}.}
  \label{fig:abu-distri}
\end{figure}

With the limitation of only having two fluids in the hydrodynamic simulations
only the energy feedback from the \nuclei{12}{C}$(\pt,\gamma)$\nuclei{13}{N}
reaction can be modeled within them. The 1D advective and diffusive
post-processing simulations that use the mixing and entrainment calibrated from
those hydro simulations can model the \ipr~by incorporating a nuclear network with
thousands of species. With the additional reactions there appear
sources of neutrons and H at the bottom of the convection zone through the
\nuclei{13}{C}$(\alpha,n)$\nuclei{16}{O} and various $(\nt, \pt)$ reactions as
seen in the left panels of \Fig{fnet}. Although not very significant, there is a
distinction between how the advective and diffusive mixing models behave with
two burning regions for a given isotope. For \code{N16}, there is the H-burning
region that ends roughly around \unit{12}{Mm}, where the $X_{\mathrm{H}}$
reaches a minimum, and then there is the H source from $(\nt, \pt)$ reactions
that occurs until the bottom CB. The advective mixing model clearly
distinguishes these two burning regions, while the diffusive mixing model smears
them, even with time steps that are needed to resolve an explicit
advective model. Of course, with larger time steps this smearing becomes much
more significant and the profiles are not properly converged. Another important
detail with the advective models is that up- and downstreams are distinct in
not only H but other isotopes as well. As the H is advected down in the
downstream it burns to produce \nuclei{13}{N} which has a half life of
approximately \unit{10}{min}. In \code{N16} the convective turn over timescale
is \unit{18}{min} leading to some of the \nuclei{13}{N} to decay as it moves to
the bottom CB and then turns over into the upstream. This can also be traced
with the \nuclei{13}{C} as it is burned while being advected by the downstream to
the bottom CB. When it eventually moves upwards in the upstream more of the
\nuclei{13}{N} decays and replenishes the burned \nuclei{13}{C}, and the streams
become homogenized.

On the right panels of \Fig{fnet} a few unstable Kr isotopes are plotted with
varying half lives. With the convective turn over timescales of \unit{18}{min}
and \unit{9}{min} for \code{N16} and \code{N17} respectively the long half life
of \nuclei{88}{Kr}, \unit{2.84}{hr}, ensures that it is well mixed as its
Damk\"ohler number ($\mathrm{Da} = \tau_{\mathrm{mix}} /
\tau_{\mathrm{nuclear}}$) is much less than 1. However for \nuclei{89}{Kr} and
\nuclei{90}{Kr}, with their half lives of \unit{3.18}{min} and \unit{32.3}{s},
their Damk\"ohler numbers become greater than 1. With these isotopes the
distinction between the up- and downstreams becomes apparent and important for
describing their distribution within the convection zone. Again, the production
of the neutrons and thus these unstable species is predominately at the bottom
CB. They are produced when the appropriate $\nt - 1$ isotope is advected down
and then captures a neutron after which it will predominately be advected away
in the upstream.

While the hydrodynamic simulations are only able to model the He-shell flash
convection and the H ingestion for roughly half a day, the latter can last in RAWDs
roughly for a month~\citep{Denissenkov:19}. A simple approximation for modeling the \ipr~over
that longer timescale is to assume that the quasi-static behavior of \code{N16}
will continue, and so the post-processing models are repeated to stretch the integration time. Likewise,
the \code{N17} simulated behavior can be repeated. The RAWD model G from~\citet{Denissenkov:19},
on which these simulations are based, had \ipr~yields that were in good
agreement with the observed elemental abundance distribution in the exemplary
carbon-enhanced metal poor (CEMP-r/s) star CS31062-050 with seemingly enhanced
\spr~and \rpr~material. There, a simpler approximation was used to compute the
\ipr~yields in which a representative time step during the H-ingestion phase was
used as a static model for \code{mppnp} with the diffusion coefficients provided
by \code{MESA} \citep{Denissenkov:19, Paxton2011}.
\Fig{abu-distri} shows the best fit in time of the decayed elemental
abundance yields from \code{mp5}, and \code{mp7}, with the corresponding
advective post-processing models \code{mp6} and \code{mp8}, compared with the surface chemical composition of
CS31062-050 from \cite{aoki:02c} and \cite{johnson:04}. Neither of the advective models has a large enough neutron
exposure at that point in time to reach the \nuclei{}{Pb} elemental abundances
observed in that star. 
The difference in the time it takes for the advective
model to reach the same \nuclei{}{Pb} abundances as in the diffusive model is approximately 7\%. This is
of the same order as the inaccuracies in the amount of H ingested and mixed, therefore
this is not due to differences in how the
\ipr~nucleosynthesis works in the advective mixing models, but is likely due to
the numerics (\Section{entrain}). From
the many RAWD models of \citet{Denissenkov:19} it is clear that the neutron
density is directly proportional to $\dot{M_{\mathrm{H}}}$ as nearly all of the
ingested H is burned via \nuclei{12}{C}$(\pt,\gamma)$\nuclei{13}{N} leading to a
neutron being released. Therefore, the best fit is reached at an earlier time
for the \code{mp7} run that is based on the \code{N17} simulations with
the higher H-entrainment rate.

The important result is that even with using the advective method that 
models mixing in RAWDs closer to what is observed in 3D hydrodynamic simulations
and uses much better time resolution we can still reproduce the observed chemical
composition of CS31062-050 well within the H-ingestion timescale of 0.087 yr estimated from
the 1D stellar evolution computations.

\section{Summary and conclusions}
\label{sec:conclusions}

In this paper, we described, formulated and applied two different numerical
methods to model the mixing within a convection zone from 3D hydrodynamic
simulations in order to compute convective-reactive \ipr~nucleosynthesis. The
stellar environment of choice was the He shell in a RAWD with a metallicity of
$\mathrm{[Fe/H]} = -2.6$ (model G of~\citet{Denissenkov:19}). One method used a
standard 1D diffusive mixing routine, while the other was a two-stream advective
mixing model that is likened to the models
of~\citet{cannon1993}~and~\citet{Henkel2017}. These models require mixing
coefficients which could be taken from estimates using MLT however instead we
constrain the mixing by running 3D hydrodynamic simulations of the RAWD. The
mixing coefficients in both models are determined directly from the data of the
3D hydrodynamic simulations.

The high resolution RAWD simulation, \code{N16}, was run for approximately 39
convective turn over times showing that its evolution over this timescale to be
approximately quasi-static. The radial velocity field at $\unit{14.5}{Mm}$ is
dominated by the large modes of $\ell = 2$ and $3$, while near the CBs the flow
is spread across many smaller modes (\FigTwo{mollweide}{power-spectrum}). The
large scale modes encounter the stiff upper boundary and begin to turn over,
increasing the tangential velocities near the boundaries (\FigTwo{rprofs}{vt-mollweide}). Using
the gradient of the tangential velocity as a condition for the CB yields a
spherically averaged boundary at $\unit{23.56}{Mm}$ at $t =
\unit{299}{minutes}$. This CB is consistent with the CB as determined by the
initial Schwarzschild boundary of the 3D simulations, which is followed in the
Lagrangian coordinates, to within a standard deviation of that averaged
boundary. This is roughly at the resolution of the \textit{briquette} data that
was used to calculate those boundaries.

The entrainment of \cldfluid{} in \code{N16} is linear in time with an
entrainment rate of $\dot{M}_{\mathrm{e}} = 7.21 \times
10^{-12}\,M_{\odot}$\,s$^{-1}$. \code{N15}'s entrainment rate increases by as
much as a factor of 3 after it experiences a global oscillation of shell H ingestion however it returns to
quasi-static burning and entrainment as shown in \Fig{entrain}. The global oscillation of shell H ingestion
instability in \code{N17} causes it go into a feedback loop of rapidly
increasing entrainment in which it does not return to quasi-static burning and
entrainment. The entrained H-rich fluid is advected along the downflows to be
burned rapidly at $\unit{14.5}{Mm}$, while the H-free fluid is advected along the
upflows. These two fluid mixtures are nearly isolated from each other as there
is a significant amount of H-free fluid with $X_{\mathrm{H}} \leq
1\times10^{-6}$ very close to the upper boundary, $\unit{23}{Mm}$, while the
spherical average is $X_{\mathrm{H}} \approx 1\times10^{-3}$ in the \ppmstar{}
simulations.

The 1D \mppnp{} post-processing simulations were first applied with only the
\nuclei{12}{C}$(\pt,\gamma)$\nuclei{13}{N} reaction which is the only reaction
included in the \ppmstar{} simulations. The instantaneous burning rate of H per
unit volume is estimated in \code{N16} and \code{mp2} which are in good
agreement as to where the bulk of the burning occurs.
The up- and downstreams of \code{mp2} have horizontal mixing timescales that
are 1-2 dex longer than the radial mixing timescales resulting in the streams
not homogenizing. Each stream has a distinct $X_{\mathrm{H}}$ radial profile
that is qualitatively similar to \code{N16}'s distribution of H-free upflows and
H-rich downflows though not of the same magnitude. The spherical average of the
two streams is consistent with the \ppmstar{} $X_{\mathrm{H}}$ profiles except
at the upper boundary where it is underestimated. The difficulty in quantifying
the convective boundary, the entrainment rate and the spatial averages done in
the computation of the \textit{briquette} data are possible sources of this
discrepancy.

With the 1D post-processing models, many more species were included for a more
elaborate burn network in order to model the \ipr~within the RAWD. The advective
mixing model sharply distinguishes the two burning regions of H, the
\nuclei{12}{C}$(\pt,\gamma)$\nuclei{13}{N} sink and the many $(\nt, \pt)$
sources at the bottom of the convection zone, while the diffusive mixing smears
these regions even with the time resolution of an explicit advective model.
However, the consequences of this subtle effect are not clear in this
application due to numerical errors, which are different for the two models,
that adjust the amount of H being ingested and accumulated in the convection
zone. This directly impacts the neutron density in each post-processing
simulation which results in the advective models requiring more simulation time
to reach the same neutron exposure as the diffusive models.

For this particular application the sharp distinction between the burning
regions of H in the advective mixing models played a minor role, which could
have been masked entirely by the numerical inaccuracies in the ingested H, in
the \ipr~yields of the RAWD. However this may not be true in the environment of
C-ingestion into a O shell of a $25\,M_{\odot}$ star~\citep{Ritter2018}. This
could be a production site for odd Z elements significant enough to influence
galactic chemical evolution of said elements~\citep{Cote2018}. The many
different burning layers of $(\gamma, \pt)$ and $(\pt, \gamma)$ that produce the
odd Z elements may be sensitive to the exact nature of the mixing of these
species throughout the burning regions to which the advective mixing model would
be well suited. The non-linearity of the \nuclei{12}{C}$+$\nuclei{12}{C} burning
could lead to interesting differences in the burning and possible
nucleosynthesis pathways of the two streams~\citep{Andrassy:19}.

The Fortran advective mixing subroutine used in this work is available at \href{https://github.com/David-Stephens/two-stream-mixing}{Github}.

\section*{Acknowledgements}

We would like to thank Marco Pignatari and Richard Stancliffe for valuable
discussions in the early phase of the project. FH acknowledges funding
from NSERC through a Discovery Grant. This research is supported by
the National Science Foundation (USA) under Grant No. PHY-1430152
(JINA Center for the Evolution of the Elements). RA, who completed
part of this work as a CITA National Fellow, acknowledges support from
the Canadian Institute for Theoretical Astrophysics and from the Klaus
Tschira Stiftung. PRW acknowledges NSF grants 1413548
and AST-1814181. The simulations were carried out on the Compute Canada supercomputer Niagara operated by SciNet at the University of Toronto and the NSF supercomputer Frontera operated by TACC at the University of Texas. 
The data analysis was carried out on the virtual research environment Astrohub \url{https://astrohub.uvic.ca} operated by the Computational Stellar Astrophysics group at the University of Victoria. Astrohub is hosted on the Compute Canada Arbutus Cloud operated by Research Computing Systems at the University of Victoria.

\section*{Data availability and software}
The data as well as all of the notebooks needed to make all 
figures of this paper are available on the virtual research platform  
\url{https://www.ppmstar.org} in the \emph{Public \& Outreach} Jupyter server that 
is found in the 
\emph{Hubs} tab. Access is granted via GitHub authentication. The notebooks are located in 
the directory \code{iRAWD-ATS\_Stephens21}
of the repository \url{https://github.com/PPMstar/PPMnotebooks}, \code{v0.9}. These notebooks are already preloaded 
 in the Jupyter session in the directory \code{PPMnotebooks/iRAWD-ATS\_Stephens21}. 

The data made available for this project includes the 3D briquette data and amounts to $>\unit{5.8}{TB}$. This data access facility is provided on a best effort basis. 

This work benefited from the use of a large amount of
free/open source software, most importantly the \code{MESA} stellar-evolution code,
\code{IPython}/\code{Jupyter} notebooks, \code{Docker}, \code{Python} and 
libraries such as \code{matplotlib}, \code{numpy}, \code{scipy}
 and \code{pyshtools},
the \code{FFmpeg} software suite, \code{Subversion} and \code{Git} 
revision control systems, and the
\code{LaTeX} document preparation system.



\bibliographystyle{mnras}
\bibliography{nugrid} 





\bsp	
\label{lastpage}
\end{document}

%% file: code.tex
\newcommand{\code}[1]{\texttt{#1}}

\newcommand{\mppnp}{\code{mppnp}} 


%% file: nuclides.tex
\newcommand{\nuclei}[2]{\ensuremath{\mathrm{^{#1}#2}}}

\newcommand{\neutron}{\ensuremath{n}}
\newcommand{\nt}{\neutron}
\newcommand{\proton}{\ensuremath{p}}
\newcommand{\pt}{\proton}

%% file: concepts.tex
\newcommand{\spr}{\mbox{$s$-process}}

\newcommand{\ipr}{\mbox{$i$-process}}
\newcommand{\iprn}{\mbox{$i$ process}}

\newcommand{\rpr}{\mbox{$r$-process}}

\newcommand{\mach}{\ensuremath{\mathcal Ma}}

%% file: units.tex
\newcommand{\unitspace}{\ensuremath{\,}}

\newcommand{\numberspace}{\ensuremath{\;}}

\newcommand{\unitstyle}[1]{\ensuremath{\mathrm{#1}}}

\newcommand{\natlog}[2]{\ensuremath{#1\times 10^{#2}}} 

\newcommand{\second}{\unitstyle{s}}   

\newcommand{\minute}{\unitstyle{min}} 

\newcommand{\unit}[2]{\ensuremath{#1\numberspace\mathrm{#2}}}

%% file: formatting.tex
\newcommand{\Tabff}[1]{{\ref{tab:#1}}}
\newcommand{\Tab}[1]{{Table~\Tabff{#1}}}

\newcommand{\pan}[1]{{\textit{#1}}}

\newcommand{\FIGFF}[2]{{\ref{fig:#2}\pan{#1}}}

\newcommand{\FIG}[2]{{Fig.~\FIGFF{#1}{#2}}}
\newcommand{\Fig}[1]{{\FIG{}{#1}}}
\newcommand{\FigTwo}[2]{{\FIGS{}{#1} and \FIGFF{}{#2}}}
\newcommand{\FIGS}[2]{{Figs.~\FIGFF{#1}{#2}}}

\newcommand{\Sectff}[1]{{\ref{sec:#1}}}

\newcommand{\Section}[1]{{Section~\Sectff{#1}}}
\newcommand{\Sections}[2]{{Sections~\Sectff{#1}~and~\Sectff{#2}}}

\newcommand{\Eqref}[1]{{\ref{eq:#1}}}

\newcommand{\eq}[1]{{equation~\Eqref{#1}}}
\newcommand{\Eq}[1]{{Equation~\Eqref{#1}}}